\documentclass[10pt,letterpaper]{article}
\usepackage[top=0.85in,left=2.75in,footskip=0.75in]{geometry}
\usepackage{arydshln}
\usepackage{changepage}
\usepackage[utf8]{inputenc}
\usepackage{textcomp,marvosym}
\usepackage{fixltx2e}
\usepackage{amsmath,amssymb}
\usepackage{cite}
\usepackage{nameref,hyperref}
\usepackage{lscape}
\usepackage[right]{lineno}

\usepackage{microtype}
\DisableLigatures[f]{encoding = *, family = * }

\usepackage{rotating}

\raggedright
\usepackage[aboveskip=1pt,labelfont=bf,labelsep=period,justification=raggedright,singlelinecheck=off]{caption}
\makeatletter
\renewcommand{\@biblabel}[1]{\quad#1.}
\makeatother

\date{}
\usepackage{lastpage,fancyhdr,graphicx}
\fancyheadoffset[L]{2.25in}
\fancyfootoffset[L]{2.25in}
\begin{document}
\vspace*{0.35in}

% Title must be 150 characters or less
\begin{flushleft}
{\Large
\textbf\newline{Finding influential spreaders from human activity beyond network location}
}
\newline
% Insert Author names, affiliations and corresponding author email.
\\
Byungjoon Min\textsuperscript{1},
Fredrik Liljeros\textsuperscript{2,3},
Hern\'an A. Makse\textsuperscript{1,*}
\\
\bf{1} Levich Institute and Physics Department, City College of New York, New York, NY, US
\\
\bf{2} Department of Sociology, Stockholm University, Stockholm, Sweden
\\
\bf{3} Institute for Futures Study, Stockholm, Sweden
\\
* E-mail: hmakse@lev.ccny.cuny.edu
\end{flushleft}

\section*{Abstract}
Most centralities proposed for identifying influential spreaders 
on social networks to either spread a message or to stop an epidemic
require the full topological information of the network on which 
spreading occurs. In practice, however, collecting all connections 
between agents in social networks can be hardly achieved. As a 
result, such metrics could be difficult to apply to real social networks.
Consequently, a new approach for identifying influential people 
without the explicit network information
is demanded in order to provide an efficient immunization or spreading strategy, in a practical sense.
In this study, we seek a possible way for finding influential spreaders by
using the social mechanisms of how social connections are formed in real networks.
We find that a reliable immunization scheme can be achieved 
by asking people how they interact with each other.
From these surveys we find that the probabilistic 
tendency to connect to a hub has the strongest
predictive power for influential spreaders among tested social mechanisms.
Our observation also suggests that people who connect different communities 
is more likely to be an influential spreader when a network has a strong modular structure.
Our finding implies that not only the effect of network location
but also the behavior of individuals is important to design optimal immunization 
or spreading schemes.
% Please keep the Author Summary between 150 and 200 words
% Use first person. PLOS ONE authors please skip this step. 
% Author Summary not valid for PLOS ONE submissions.   
\section*{Author Summary}
%Design of efficient immunization strategies is important 
%to lower the threshold of immunization against an infectious disease.
%For this, the most influential spreaders should be immunized 
%with the highest priority in order to reduce epidemics.
%Finding influential spreaders is also important for viral marketing 
%campaign to search efficient strategy of spreading via targeting superspreaders.
%Here, we propose an approach for identifying influential people based on 
%social mechanisms of individuals, that is how connections are formed in real-world social networks.
%Since these properties can be obtained by using a survey conducted for a population,
%it is easy to apply for real spreading.
%We find that the social mechanism of connecting a hub 
%can be reliable predictors for influential spreaders in future 
%epidemics or viral marketing campaigns.
%From the analysis of the microscopic link formation, 
%we find the effect of social mechanisms on spreading processes.

\section*{Introduction}
Identifying influential spreaders on social networks is crucial for 
its practical application in real-world epidemic and information spreading 
\cite{cohen,chen,holme_immun,domingos,domingos2}.
For instance, superspreaders need to be immunized 
with the highest priority in order to prevent the pandemic of 
an infectious disease \cite{albert,holme_attack,kitsak,castellano,pei}.
They are also important for spreading of information in viral marketing 
\cite{domingos,domingos2,kempe,kempe2}.
To this end, several predictors for influential spreaders based on 
the topological property of complex networks, including high degree \cite{albert,havlin}
$k$-core \cite{kitsak,dorogovtsev,carmi}, 
betweenness centrality \cite{freeman}, PageRank \cite{page}, and many others \cite{pei_makse} were 
tested for identifying influential spreaders \cite{kitsak,castellano,pei}.

Most studies, however, have overlooked how to apply to the real-world social systems 
which is a serious problem in a practical sense.
Most proposed centralities except the degree, which is a local centrality, 
require the information of the whole network structure.
But collecting this information is nearly impracticable in real social systems. 
Specifically, gathering information of relationships among individuals 
is inevitably incomplete and erroneous \cite{lee}, since it cannot but be conducted 
for a partial sample of a whole population.
Thus, searching for the influential spreaders with these centralities 
may not be plausible for real-world spreading phenomena.
On the other hand, if whole connections in a network are accessible, direct 
measuring for the influence of a single node is possible by using model simulation 
on the network, which obviates the need for predicting influential spreaders.
Consequently, in reality most predictors proposed for an influential 
spreader are either inapplicable or unnecessary.

Thus more realistic approaches based on the characteristics of people 
such as their behaviors are demanded for predicting influential 
people without the explicit information of network structure.
The benefit of this method is an easy applicability for any kinds of social networks
since one can obtain the probabilistic actions of agents by using a survey conducted from a population. 
Through a survey, we can estimate the probabilistic tendency of how connections are established for 
each individual, for instance, how probable is to make a new friend by introduction from another friend
or the frequency to make new friends from different groups.
We find that these human actions have a large influence on the
subsequent spreading of information and therefore 
can be a reliable predictor of the node's importance in a future epidemic
or in a viral marketing campaign via targeting people identified by
their probabilistic actions.
In addition, such ranking obtained from surveys can also apply to the situations when
the information for only some people is accessible.

The social mechanisms of link formation driving evolution of networks
have been studied for a long time in order to explain 
and predict complex phenomena in society.
A number of social mechanisms for connection establishments 
have been proposed in sociology \cite{contractor,monge}.
Thanks to the detailed records in online social networks that captures the action of every individual,
it is now possible to quantify the frequency of occurrence of 
different types of mechanisms by directly observing social interactions \cite{gallos}.
Thus, recently, the frequencies of the social mechanisms for each person in a social network 
have been revealed from the full log of the activity in online social networks \cite{gallos}.

In this paper, we propose an approach to identify influential spreaders based on 
surveys on human behavior and social mechanisms 
that can be given to a population without the explicit information of networks. 
We decode the relation between people's characteristics that can be obtained by a survey
and their influence in spreading using the real-world datasets that
contain the full information of network evolution.
Through the analysis of large-scale evolving networks,
we identify the effect of the microscopic link formation on macroscopic consequences in spreading.
We find that the interaction to connecting a hub 
can facilitate epidemic spreading and thus can be a reliable
predictor of people's importance in future epidemics or viral marketing campaigns.
We also find that people with high frequency to connect different communities 
are more likely to be an influential spreader for the case when a network is composed of strongly connected modules.
This research represents much to practical implication,
since our finding can be adopted in reality requiring only the tendency of individuals' behaviors.
Furthermore, our results provide a guideline for behavior to the public,
about how to behave at the beginning stage of epidemic.

% You may title this section "Methods" or "Models". 
% "Models" is not a valid title for PLoS ONE authors. However, PLoS ONE
% authors may use "Analysis" 
\section*{Materials and Methods}
\subsection*{Social mechanism}
In this paper, the social mechanisms are referred to as 
the probabilistic tendency of each kind of interaction 
among people in a given social network.
The social mechanisms do not directly mean the motivation behind the link creation
because several different mechanisms may result in the same type of link formation
and link formation may not be motivated by only the structure \cite{gallos}.
In addition, these mechanisms are not complementary one another, because 
a link can be established by multiple different mechanisms.
For instance, a newly created link can appear following balance and exchange interactions at the same time.

We use four classes of social mechanisms underlying the link 
creation on a network based on the multitheoretical multilevel 
formalism \cite{contractor} proposed in sociology: 
(1) Exchange interaction corresponds to a newly form reciprocal link meaning that 
a new link is established in the opposite direction of an existing link.
(2) Balance interaction corresponds to a newly form tie that closes a triangle by a directed edge.
(3) Collective action (or preferential attachment \cite{barabasi}) corresponds to
a link that connects with well-connected people.
To be specific, in this study, we measure the extent of the collective action of each link 
as a continuous value using the cumulative probability $F(k_i)$ 
of the excess degree distribution for a newly connected neighbor $i$. 
Here, $F(k)=\sum_{k_j<k} k_j q(k_j)/\langle k\rangle$ where $q(k)$ is the degree distribution of a network
and $\langle k\rangle$ represents the average degree of a network.
(4) Structural hole interaction considers a newly created link that
connects two different modules (communities).
Community structure is identified by the local version of 
link community detection method \cite{ahn} when a new link is established
[see detailed in Text~S2].

These social mechanisms are assigned on an evolving network 
at the moment when the link is newly added following the analysis developed in \cite{gallos}.
While constructing the evolving network
by adding the new connection in sequential order,
we characterize each connection to the corresponding social mechanisms 
based on a network configuration at the given moment.
After all links are formed, the frequencies of social mechanisms of the origin node, $i$,
$a_i^{\rm exc}$, $a_i^{\rm bal}$, $a_i^{\rm ca}$, and $a_i^{\rm sh}$, 
where $i$ is node index, are defined as the number of neighbors 
that were connected by the corresponding mechanism, respectively, 
exchange, balance, collective action, and structural hole
(the sum of the extent for the collective action of all connected nodes)
normalized by the total number of neighbors.
To be specific, the frequency $a^{\alpha}_i$ of social mechanism $\alpha$ for node $i$
is defined as $a^{\alpha}_i=\frac{n^{\alpha}_i}{k^{out}_i}$,
where $n^{\alpha}_i$ is the number of links formed corresponding to $\alpha$ social mechanism
and $k_i^{out}$ is the outdegree (the total number of new connections).
Therefore, each variable ranges from zero to unity, and
as $a_i$ increases, the corresponding social interaction is more frequent.

We stress here that the extent of social mechanisms of link creation for each individual
can be estimated in a real setting by the surveys given to the population. 
For instance, one first could ask people to list their contacts
and then as a second stage ask questions about each contact \cite{fridlund}.
For example, we could ask questions like,
(exchange) did the person contact you first?, 
(balance) did the person have common friends with you when you contacted him/her?, 
(collective action) did the person have a lot of contacts when you contacted him/her?,
(structural hole) did person belong to another group than you when you contacted him/her?
Therefore, an estimate of $a_i$ for each individual can be obtained from the surveys conducted for the population.
On the contrary, most centralities including $k$-shell index \cite{kitsak},
betweenness centrality \cite{freeman}, and PageRank \cite{page} cannot be obtained by this way 
since they require global network information.

\subsection*{Data sets}
We examine two social networks of Internet dating services in Sweden \cite{gallos,holme_pok}
and the forum of internet-mediated prostitution in Brazil \cite{rocha_pro}.
These social networks represent potential pathways for epidemic spreading
including sexually transmitted diseases. 
We use the data of the largest site {\it qx.se} for Nordic homosexual, bisexual, 
and transgender people in 2006 (QX).
Actions of every individual in the community, including adding an individual to the favorite list 
and guestbook signing, were recorded for two months starting from Nov. 2005.
We use adding favorite lists (QXF) and signing guestbook lists (QXG) among many activities.
We also analyze pussokram {\it pussokram.com} dataset (POK) \cite{holme_pok}, which was a Swedish online dating 
site for friendship including flirting and non-romantic relations.
The data contains a full log for 512 days starting from the day when the 
community was created in 2011.
The POK network that we use in this study consists of message senders, receiver, 
and the timing of interactions in the community.
Internet-mediated prostitution data (PRO) \cite{rocha_pro} comes from Brazilian online forum 
where sex-buyers evaluate prostitutes.
We construct the PRO network by connecting sex-sellers with buyers.
Since the PRO network is an undirected and bipartite graph, 
the exchange and balance interactions are not defined.
In order to investigate the problem of identifying influential spreaders of information,
we study the citation network in the posts of an online 
network service, {\it livejournal.com} (LJ), for information spreading on social networks \cite{pei}.
One should note that the QX has already a large part of network 
(85 and 87 $\%$ for the QXF and QXG, respectively)
whereas the others starts at time $t=0$.
Table~1 gives the basic information of the datasets.

\begin{table}[t]
\caption{
{\bf Properties of real-world networks used in this study. }
}
\begin{tabular}{|l|l|l|l|l|}
\hline
Network & Name  & Number of nodes & $\langle k \rangle$ & Modularity \cite{clauset} \\
\hline
QX.com favorite & QXF & 80,407 & 13.07 & 0.4060  \\
\hline
QX.com guestbook & QXG & 59,854 & 7.10 & 0.3893  \\
\hline
POK.com & POK & 29,242 & 5.95 & 0.3992  \\
\hline
Livejournal.com & LJ & 315,936 & 3.56 & 0.6578  \\
\hline
Prostitution & PRO & 16,729 & 4.67 & 0.6294  \\
\hline
\end{tabular}
\begin{flushleft} 
	$\langle k \rangle$ is the average degree of the network.
	We use the fast-greedy community detection algorithm \cite{clauset} for measuring modularity.
\end{flushleft}
\label{table1}
\end{table}

We can reconstruct the evolving connection of networks, following 
the precise timing when a tie has been established, in contrast to
the observation of static snapshots of networks.
In our datasets, we can observe every evolution of social networks
with the time stamp of link creations.
We stress here that the precise information of temporal evolution is essential 
to identify the social mechanisms for each link.
The social mechanisms should be defined at the moment when a new link established \cite{gallos}.
Accumulated static networks do not keep the order of time that links established
and thus are misleading about the social interactions.
In this regard, our datasets containing the full log of network evolution
allow us to define social mechanisms properly.

\subsection*{Influential spreader}
In order to assess the influence of people for epidemic spreading, 
we use the epidemic size $M_i$ originating from a seed $i$ in 
the susceptible-infected-recovered (SIR) model on the finally accumulated network \cite{kitsak}.
The SIR model has been used to describe infectious disease for a long time \cite{anderson_may}.
At the same time, the SIR model is a plausible model of information spreading \cite{kitsak}.
In the SIR model, each node can be in one of three states, susceptible, infected,
or recovered (or removed).
Initially, all nodes are in the susceptible state except for a single node in the infected state.
At each time, the infected node spread a disease/information to a susceptible neighbor with 
infection probability $\beta$.
At the steady state, we measure $M_i$ as the fraction of finally infected nodes.
We define a node with high $M_i$ as highly influential.

We choose the infection probability $\beta$ to be a value covering 
a small part of a network, $\beta \gtrsim \beta_c$ 
where $\beta_c$ is the epidemic threshold for percolation \cite{anderson_may,newman_epi}.
When $\beta \gg \beta_c$, all seed produces
similar epidemic size because spreading can cover almost all network 
regardless of where it originated from \cite{pei_plos}.

% Results and Discussion can be combined.
\section*{Results}

\subsection*{Predictor for influential spreaders based on human activity.}
We recreate the entire network by adding all links in the order of time that they were established.
In order to assess systematically the relation of the epidemic influence $M_i$ 
with the social mechanisms as well as topological metrics, 
we use multilinear regression analysis \cite{gujarati}
with the following model (Tables~S1-S5):
\begin{equation}
M_i=c_0+c_1 a^{\rm exc}_i +c_2 a^{\rm bal}_i +c_3 a^{\rm ca}_i + c_4 a^{\rm sh}_i + c_5 k_i 
	+c_6 k^{\rm sh}_i +c_7 k^{\rm 2sum}_i +c_8 k^{\rm sum}_i+\epsilon.
\end{equation}
Here, $k_i$ is the degree of node $i$, $k_i^{\rm sh}$ is the $k$-shell index \cite{kitsak} (Text~S1).
$k_i^{\rm sum}$ is the sum of degree of the nearest neighbors $k_i^{\rm sum}=\sum_{j\in V(i)} k_j$ where $V(i)$ is 
the set of node $i$'s neighbors \cite{pei},
$k_i^{\rm 2sum}$ is the sum of degrees of the next-nearest neighbors,
$k^{\rm 2sum}_i=\sum_{j\in V_2(i)} k_j$ where $V_2(i)$ is 
the set of neighbors of node $i$'s neighbors \cite{pei}, and $\epsilon$ is the error term.
We introduce the topological metrics, 
since we are interested in how much information we captured
using the social mechanisms tendencies $\{a_i^{\rm exc},a_i^{\rm bal},a_i^{\rm ca},a_i^{\rm sh}\}$
in comparison with the more common topological measurements,
$\{k_i$, $k_i^{\rm sh}$, $k_i^{\rm sum}, k_i^{\rm 2sum} \}$.
In order to avoid biased observation due to the large fluctuation in the small degree region,
we exclude the data of people with degree less than three from our analysis.

\begin{figure}
\begin{center}
\includegraphics[width=\linewidth]{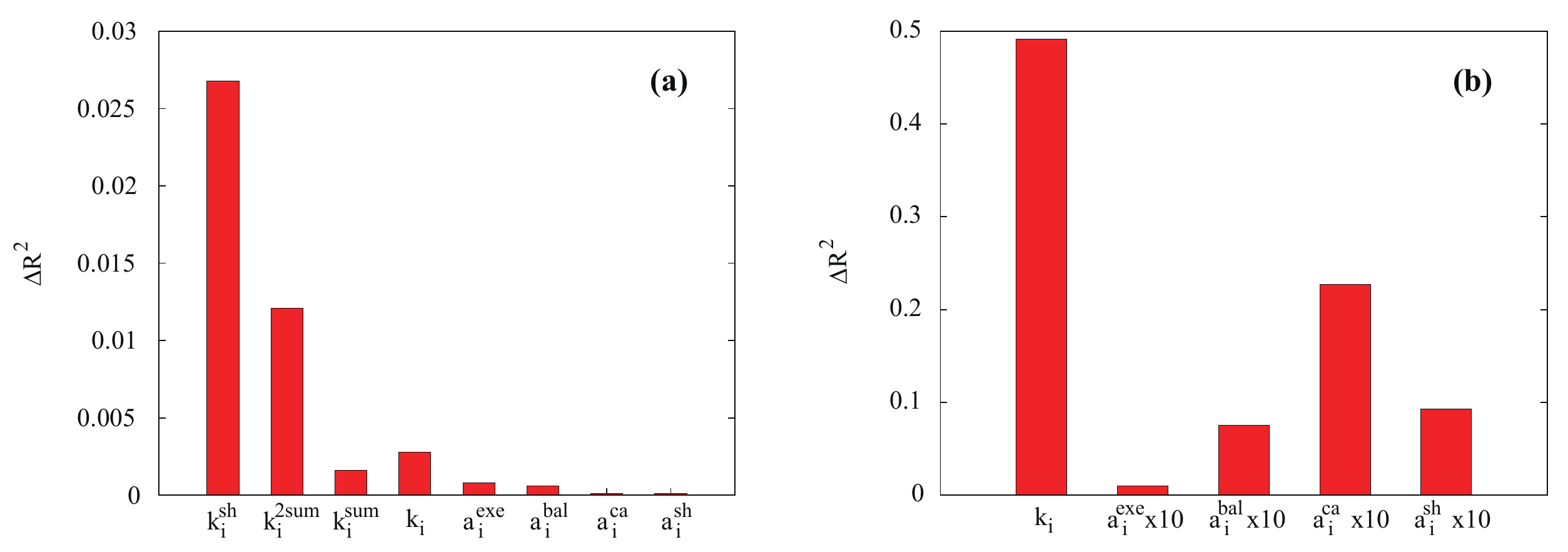}
\caption{{\bf The difference $\Delta R^2(x)$ of the coefficient of determination when a 
	variable $x$ is excluded in regression analysis of (a) Eq.~(1) and (b) Eq.~(2) in QXF network.}
	(a) $k$-shell index shows the largest drop of $R^2$, showing the strongest 
	predictive power for influential spreaders.
	However, $k$-shell index is difficult to obtain since it
	requires the full topological information of the network.
	Although the degree and social mechanisms $a^{\alpha}$ show
	smaller predictive power than $k$-shell index,
	they can be easily obtained from surveys and have much 
	implication in a real setting.
	(b) The degree shows the largest difference among the degree and social mechanisms 
	that can be obtained from surveys.
	Next, among the social mechanisms, collective action shows the largest 
	drop of $R^2$. Thus, collective action is a more reliable predictor than the others 
	from the human behavioral point of view.
}
\end{center}
\end{figure}

%\begin{figure}
%\begin{center}
%\includegraphics[width=\linewidth]{./Fig1.pdf}
%\caption{{\bf The difference $\Delta R^2(x)$ of the coefficient of determination when a 
%	variable $x$ is excluded in regression analysis of (a) Eq.~(1) and (b) Eq.~(2) in QXF network.}
%	(a) $k$-shell index shows the largest drop of $R^2$, showing the strongest 
%	predictive power for influential spreaders.
%	However, $k$-shell index is difficult to obtain since it
%	requires the full topological information of the network.
%	Although the degree and social mechanisms $a^{\alpha}$ show
%	smaller predictive power than $k$-shell index,
%	they can be easily obtained from surveys and have much 
%	implication in a real setting.
%	(b) The degree shows the largest difference among the degree and social mechanisms 
%	that can be obtained from surveys.
%	Next, among the social mechanisms, collective action shows the largest 
%	drop of $R^2$. Thus, collective action is a more reliable predictor than the others 
%	from the human behavioral point of view.
%}
%\label{Figure_label}
%\end{center}
%\end{figure}

The $k$-shell index and its local proxy $k^{\rm sum}$ and $k^{\rm 2sum}$
have been regarded as an efficient topological predictor for influential spreaders \cite{kitsak,pei}.
In agreement with these previous studies, we find that $k^{\rm sh}$ can capture 
most of the fluctuation in the epidemic size for the datasets.
To quantify the effect of each variable, we measure the difference $\Delta R^2(x)$ 
of the coefficient of determination when a variable $x$ is excluded.
In Fig.~1, the difference $\Delta R^2(k^{\rm ks})$ of the coefficient of determination
is the largest when $k^{\rm ks}$ is excluded from Eq.~(1), which confirms the importance of $k^{\rm ks}$.
In addition, more than $82.3$ $\%$ of the fluctuations can be explained 
by solely the $k$-shell index for the QXF network (Table~S1).
For the QXG, POK, LJ, POK networks, we also find the similar 
trend as the QXF (Tables~S2-S5).
However, being a global quantity, the $k$-shell index can be difficult to obtain as discussed above.
Therefore, $k^{\rm sh}$ has the limitation to apply for real social systems 
despite its strong correlation with the spreading influence.
$k^{\rm sum}$ or $k^{\rm 2sum}$ also captures a huge part of the variance in the data.
While these are a local measurement, they still can be difficult to obtain 
because they require the exact number of friends of friends at the time when epidemic occurs \cite{pei}.

The degree $k$ is not behavioral but
in contrary to $k^{\rm sh}$, $k^{\rm 2sum}$, and $k^{\rm sum}$, 
the degree $k$ can be estimated by a survey to individuals 
by a simple question: how many friends do you have?
Therefore, even if we cannot conceive the structure of network,
for many cases, we can access the information of the degree 
together with the other social mechanisms, $a_i^{\alpha}$.
Next we are interested in the case where the topological
location such as $k$-shell cannot be obtained for the reasons explained above. 
Therefore, we regress the data of $M_i$ with the variables which 
can be easily obtained by surveys using the following model,
where $k^{\rm sh}$, $k^{\rm sum}$, and $k^{\rm 2sum}$ are excluded:
\begin{equation}
M_i=c_0+c_1 a^{\rm exc}_i +c_2 a^{\rm bal}_i +c_3 a^{\rm ca}_i + c_4 a^{\rm sh}_i + c_5 k_i+\epsilon.
\end{equation}
When we consider Eq.~(2), we can explain 63 $\%$ of 
the variance for the QXF network (Table~S6),
demonstrating that with only surveys we
can capture extremely high amount of the variance.
The all variables in Eq.~(2) can be easily obtained 
from surveys, suggesting that we can rely on surveys for
optimally immunization or viral marketing.

Using Eq.~(2), we find that the degree is the most reliable predictor 
for the influential spreaders among the degree and $a_i^{\alpha}$.
When the degree is excluded from Eq.~(2), the coefficient of determination $R^2$
drops $0.49$ from $0.62$, showing the largest difference (Fig.~1b).
Since the degree represents the number of the transmission channels 
for a seed, the degree can play an important role in epidemic 
spreading on networks especially at the beginning stage of outbreak \cite{kitsak,pastor}.
When compared with the topological location of the people given by $k$-shell, 
we find that the degree alone can explain 58 $\%$ of the variance, which 
compared to the value of $k$-shell ($R^2=0.82$), indicating that 
the degree is a worse predictor than $k$-shell in agreement 
with \cite{kitsak}. In a real setting, however, the local degree
can have more implication than $k$-shell because it can be easily 
obtained from surveys.

\begin{figure}
\begin{center}
\includegraphics[width=\linewidth]{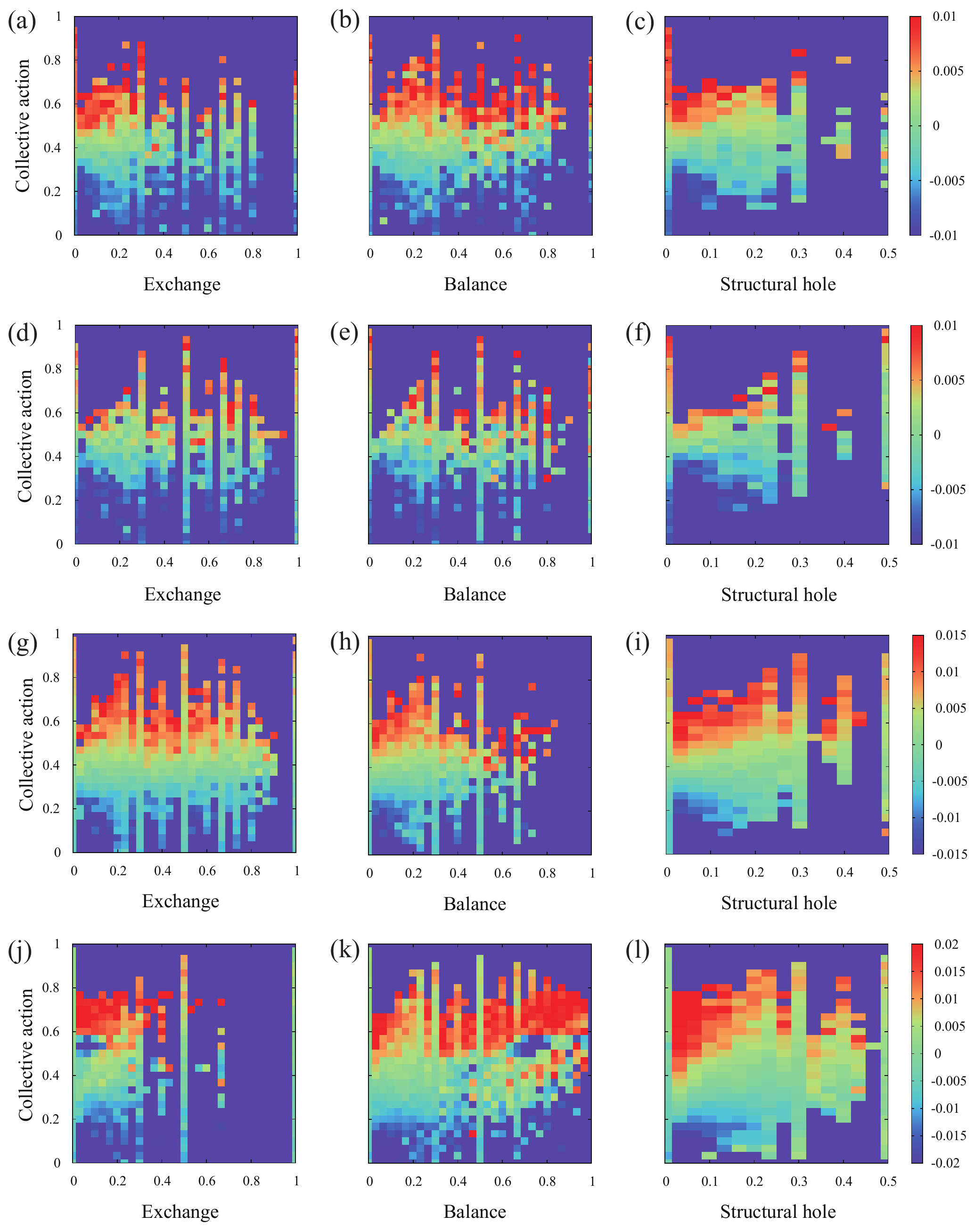}
\caption{{\bf Collective action predicts influential spreaders more reliably
than other social mechanisms.}
When spreading originates in people with $(a^{\alpha},a^{\rm ca})$, the relative 
epidemic size $M(a^{\alpha},a^{\rm ca})$ for the QXF with (a) $a^{\rm exc}$, 
(b) $a^{\rm bal}$, and (c) $a^{\rm sh}$, (d-f) QXG, (g-i) POK, and 
(j-l) LJ networks.
Collective action $a^{\rm ca}$ predicts the epidemic 
influence more reliably than the other social interactions when 
we compare for people with the same degree. 
}
\end{center}
\end{figure}

%\begin{figure}
%\begin{center}
%\includegraphics[width=\linewidth]{./Fig2.pdf}
%\caption{{\bf Collective action predicts influential spreaders more reliably
%than other social mechanisms.}
%When spreading originates in people with $(a^{\alpha},a^{\rm ca})$, the relative 
%epidemic size $M(a^{\alpha},a^{\rm ca})$ for the QXF with (a) $a^{\rm exc}$, 
%(b) $a^{\rm bal}$, and (c) $a^{\rm sh}$, (d-f) QXG, (g-i) POK, and 
%(j-l) LJ networks.
%Collective action $a^{\rm ca}$ predicts the epidemic 
%influence more reliably than the other social interactions when 
%we compare for people with the same degree. 
%}
%\label{Figure_label}
%\end{center}
%\end{figure}

Next, we are interested in what social mechanisms $a_i^{\alpha}$ 
are more important for spreading besides the local degree.
This is not only important for optimal immunization and information 
spreading but also for education of the population to avoid
certain behaviors that could spread diseases to huge population.
In order to examine the effect of the social mechanisms clearly,
we study the deviation of the epidemic size $\Delta M_i$ from 
the average epidemic size for people with the same degree by following
\begin{equation}
\Delta M_i = M_i - \frac{\sum_j \delta_{k_i,k_j} M_j}{\sum_j \delta_{k_i,k_j}},
\end{equation}
where $\delta_{i,j}$ represents the Kronecker delta such that the 
function is $1$ if the variables are equal and $0$ otherwise.
$\Delta M_i$ quantifies the impact of the social mechanisms after 
removing the effect induced by the degree, thus, more clearly 
identify the important social mechanism for spreading for 
people with the same degree.

To compare the influence of each social mechanisms in the spreading 
process, we study the average size $\Delta M$ infected in an epidemic 
originating at people $i$ with a given ($a^{\rm exc}_i,a^{\rm bal}_i,a^{\rm ca}_i,a^{\rm sh}_i$).
The average infected population over all the origins with the same 
pair of ($a^{\alpha},a^{\beta}$) is 
\begin{equation}
\Delta M = \sum_{i \in W(a^{\alpha},a^{\beta})} \frac{\Delta M_i}{N(a^{\alpha},a^{\beta})},
\end{equation}
where $W(a^{\alpha},a^{\beta})$ is the union of all nodes 
with $(a^{\alpha},a^{\beta})$ and $N(a^{\alpha},a^{\beta})$ is the 
number of nodes with $(a^{\alpha},a^{\beta})$.
In Fig.~2, we find that $\Delta M$ increases with increasing $a^{\rm ca}$ 
regardless with the other social mechanisms for all tested networks.
This clear pattern suggests that $a^{\rm ca}$ predicts the epidemic 
influence more reliably than the other social interactions when 
we compare for people with the same degree. 

The regression analysis of Eq.~(2) also supports the importance of 
the collective action. 
When we remove $a_i^{\rm ca}$ from Eq.~(2), 
the difference $\Delta R^2$ of the coefficient of determination
is the largest, which confirms the importance of collective action.
Since people with high collective action are more likely 
to have many next nearest neighbors.
they have high chance to develop larger epidemic outbreaks.
On the contrary, people with less collective action,
is likely to be located at the periphery of a 
network leading to a small impact in the spreading.
Thus, the collective action is a reliable predictor from the 
human behavioral point of view when we factor out the popularity.

\subsection*{Strength of weak ties and community structure.}
So far, we search the most influential spreaders based on social 
mechanisms, $a_i^{\alpha}$ which can be obtained by surveys.
In sociology, a long-standing hypothesis for influential spreaders
is the strength of weak ties \cite{granovetter}.
According to the hypothesis, weak ties which bridge between two 
densely connected modules formed by strong ties play an important role 
especially in the job changing in labor market \cite{granovetter,montgomery},
mobile communication networks \cite{onnela}, as well as brain \cite{sigman}.
While this hypothesis may seem counter-intuitive,
for a perspective of information spreading, the weak ties is 
more likely to be a source of fresh information, so weak ties can have
a stronger effect than strong ties.

\begin{figure}
\begin{center}
\includegraphics[width=0.7\linewidth]{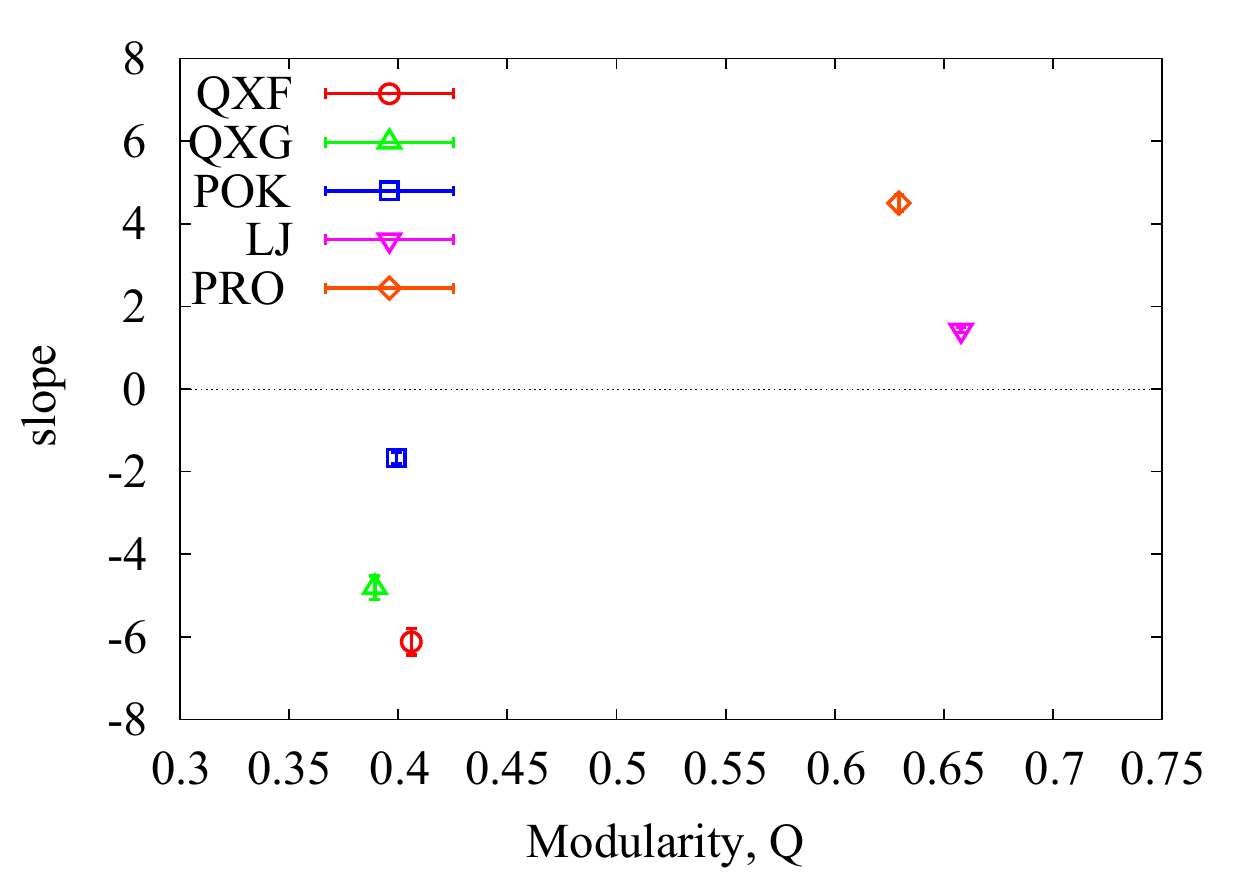}
\caption{
{\bf The effect of weak ties on spreading for different networks with diverse modularity.}
The panel shows the slope of the frequency of structural hole with respect to
epidemic influence $M_i$ in regression analysis as a function of
modularity of a underlying network.
For networks with highly modular structure such as LJ and PRO,
the frequency of structural hole is positively correlated with $M_i$.
}
\end{center}
\end{figure}

%\begin{figure}
%\begin{center}
%\includegraphics[width=0.7\linewidth]{./Fig3.pdf}
%\caption{
%{\bf The effect of weak ties on spreading for different networks with diverse modularity.}
%The panel shows the slope of the frequency of structural hole with respect to
%epidemic influence $M_i$ in regression analysis as a function of
%modularity of a underlying network.
%For networks with highly modular structure such as LJ and PRO,
%the frequency of structural hole is positively correlated with $M_i$.
%}
%\label{Figure_label}
%\end{center}
%\end{figure}

In this section, we test the weak tie hypothesis by observing the 
evolution of link formation in a large scale real-world network.
We define weak connection as a link bridging two different communities 
at the time when a new link is formed, called structural hole.
If weak ties play an important role in spreading processes as the 
hypothesis of weak ties, people with high probability of structural hole interactions 
is more likely to have influence in spreading.
In order to test the effect of weak ties (structural hole),
we regress the data of $M_i$ with the variables of social mechanisms $a_i^{\alpha}$
using the following model,
where the network properties $k^{\rm sh}$, $k^{\rm sum}$, $k^{\rm 2sum}$, and $k$ are excluded:
\begin{equation}
M_i=c_0+c_1 a^{\rm exc}_i +c_2 a^{\rm bal}_i +c_3 a^{\rm ca}_i + c_4 a^{\rm sh}_i +\epsilon.
\end{equation}
The degree $k$ is also excluded in order to focus on the effect 
of behavioral factors on spreading.
%Note that the regression analysis of Eq.~(2) including $k$
%results in qualitatively the same consequences with that of Eq.~(5) (Tables~S6-S10).

From the regression analysis, we confirm that people with high 
frequency of structural hole interaction is more likely to be 
an influential spreaders on LJ and PRO networks as the weak tie hypothesis.
In LJ and PRO networks, the frequency of structural hole $a^{\rm sh}$ is 
positively related with the spreading influence $M_i$ with extremely small
p-value (Fig.~3 and Tables~S9 and S10). 
However, this pattern does not hold for all social networks that we tested.
For QXF, QXG, and POK networks, $a_i^{\rm sh}$ is negatively correlated 
with $M_i$ in contrary to the weak tie hypothesis (Fig.~3 and Tables~S6-S8).
This result suggests that the weak tie hypothesis may not be generically
valid for all social networks. 

The validity of the weak tie hypothesis can rely on
the underlying network where spreading occurs.
People with high frequency of structural hole interactions 
potentially spreads different communities all together.
Therefore, if an underlying network of spreading has clear 
module structure, the effect of weak ties is significant \cite{masuda}.
However, when community structure is less clear
the role of weak ties in spreading can be weakened.
In order to check this prediction, we compare the modularity
of networks \cite{modularity} and the effect of weak ties (Fig.~3).
When a network has strong community structure such as LJ and PRO
whose modularity is $0.658$ and $0.629$, respectively,
the frequency of structural hole is positively correlated with $M_i$.
Therefore, the structural hole mechanisms can enhance 
the epidemic influence for networks with strong modular structure as 
the weak tie hypothesis.
However, the weak tie hypothesis is not valid for networks 
with less clear module structure.
For instance, the QXF, QXG, and POK networks showing less 
modularity around $0.4$, $a_i^{\rm sh}$ play a minor role in 
spreading and negatively correlated with $M_i$ (Fig.~3 and Tables~S6-S8).
If the modular structure is not significant, the weak ties are 
not clearly defined, leading to decrease of the effect of weak ties.
Thus, the weak tie hypothesis is expected to be 
valid for strong module structure not universally for all social networks.
In conclusion, people who connect different communities can be suspected as an influential people
when an underlying network is composed of strong modular structure,

\section*{Discussion}
So far, most studies of spreading on complex networks have assumed that 
a network structure is known.
This means that full information on any people on who is connected with whom is required,
which may not be obtained in real settings.
In agreement with the previous studies, we find that when the information of global structure of social networks
is available, it is beneficial for identifying influential spreaders in an epidemic model
capturing up to 90 $\%$ of the variance with simple variables with the $k$-shell \cite{kitsak,pei}.
In reality, however, it is difficult to gather the complete sets of interactions among people.
Therefore, all the previous method for the influential spreaders based on the network topology
could be impractical.
Searching for influential spreaders without the information of a network
is essential in order to prevent the global pandemic and minimize the cost for immunization.

Thus, we proposed a possible strategy for identifying the influential spreaders 
by using  characteristics of people's behavior underlying the evolution of social networks.
Our finding provides several pragmatic lessons for the efficient immunization strategy
as well efficient information spreading campaigns.
First, in the absence of $k$-shell, the degree is the first local quantity that
can be used to predict the influential spreaders. 
From the behavioral variables quantifying the social mechanisms $a_i^{\alpha}$,
collective action gives a complementary information to the degree,
so it is suitable for a strong indicator for influential spreaders 
when comparing among people with the same degree. 
Also, a person with a high tendency to 
connect two different groups via weak ties can also be suspected as a influential spreader
when the network has a strong modular structure.
Our analysis provide not only an applicable identifying scheme of influential spreader
based on surveys but also a guideline for activity to the public,
about how to behave when epidemic occurs.
For instance, during the beginning stage of epidemic,
one need to avoid meeting popular people or people belonging to 
a different group that could spread diseases to huge population.

\section*{Supporting Information}

\subsection*{S1 Table}
\label{S1_Table}
{\bf Multilinear regression for the QXF networks with Eq.~(1)}
$M_i=c_0+c_1 a^{exc}_i +c_2 a^{bal}_i +c_3 a^{ca}_i + c_4 a^{sh}_i + c_5 k_i +c_6 k^{sh}_i +c_7 k^{2sum}_i +c_8 k^{sum}_i+\epsilon.$

\subsection*{S2 Table}
\label{S2_Table}
{\bf Multilinear regression for the QXG networks with Eq.~(1).}
$M_i=c_0+c_1 a^{exc}_i +c_2 a^{bal}_i +c_3 a^{ca}_i + c_4 a^{sh}_i + c_5 k_i +c_6 k^{sh}_i +c_7 k^{2sum}_i +c_8 k^{sum}_i+\epsilon.$

\subsection*{S3 Table}
\label{S3_Table}
{\bf Multilinear regression for the POK networks with Eq.~(1).}
$M_i=c_0+c_1 a^{exc}_i +c_2 a^{bal}_i +c_3 a^{ca}_i + c_4 a^{sh}_i + c_5 k_i +c_6 k^{sh}_i +c_7 k^{2sum}_i +c_8 k^{sum}_i+\epsilon.$

\subsection*{S4 Table}
\label{S4_Table}
{\bf Multilinear regression for the LJ networks with Eq.~(1).}
$M_i=c_0+c_1 a^{exc}_i +c_2 a^{bal}_i +c_3 a^{ca}_i + c_4 a^{sh}_i + c_5 k_i +c_6 k^{sh}_i +c_7 k^{2sum}_i +c_8 k^{sum}_i+\epsilon.$

\subsection*{S5 Table}
\label{S5_Table}
{\bf Multilinear regression for the PRO networks with Eq.~(1).}
$M_i=c_0+c_1 a^{exc}_i +c_2 a^{bal}_i +c_3 a^{ca}_i + c_4 a^{sh}_i + c_5 k_i +c_6 k^{sh}_i +c_7 k^{2sum}_i +c_8 k^{sum}_i+\epsilon.$

\subsection*{S6 Table}
\label{S6_Table}
{\bf Multilinear regression for the QXF networks with Eqs.~(2) and (5).}
$M_i=c_0+c_1 a^{exc}_i +c_2 a^{bal}_i +c_3 a^{ca}_i + c_4 a^{sh}_i + c_5 k_i+\epsilon.$

\subsection*{S7 Table}
\label{S7_Table}
{\bf Multilinear regression for the QXG networks with Eqs.~(2) and (5).}
$M_i=c_0+c_1 a^{exc}_i +c_2 a^{bal}_i +c_3 a^{ca}_i + c_4 a^{sh}_i + c_5 k_i+\epsilon.$

\subsection*{S8 Table}
\label{S8_Table}
{\bf Multilinear regression for the POK networks with Eqs.~(2) and (5).}
$M_i=c_0+c_1 a^{exc}_i +c_2 a^{bal}_i +c_3 a^{ca}_i + c_4 a^{sh}_i + c_5 k_i+\epsilon.$

\subsection*{S9 Table}
\label{S9_Table}
{\bf Multilinear regression for the LJ networks with Eqs.~(2) and (5).}
$M_i=c_0+c_1 a^{exc}_i +c_2 a^{bal}_i +c_3 a^{ca}_i + c_4 a^{sh}_i + c_5 k_i+\epsilon.$

\subsection*{S10 Table}
\label{S10_Table}
{\bf Multilinear regression for the PRO networks with Eqs.~(2) and (5).}
$M_i=c_0+c_1 a^{exc}_i +c_2 a^{bal}_i +c_3 a^{ca}_i + c_4 a^{sh}_i + c_5 k_i+\epsilon.$

%\subsection*{S1 Figure}
%\label{S1_Figure}
%{\bf Scatter plots of the epidemic size as a function of the degree of node for the QXG, POK, and LJ networks.} 
%Color of symbols of each column represents the extent of social interactions for the 
%exchange, balance, collective action, and structural hole.
%The black line represents the average epidemic size for 
%the randomized networks conserving the degree distribution, obtained analytically.
%The collective action mechanism indicates the strong gradation horizontally.
%The nodes with high balance tendency are located in the high influential region.

%\subsection*{S2 Figure}
%\label{S2_Figure}
%{\bf The average epidemic size $M(a,a^{bal})$}
%The average size of epidemic $M(a,a^{bal})$ when spreading originates in nodes with 
%$(a,a^{bal}$ for the QXF with (a) $a^{ex}$, (b) $a^{ca}$, and (c) $a^{sh}$,
%(d-f) QXG, (g-i) POK, and (j-l) LJ networks.
%The average size of epidemic increases with increasing 
%the extent of the balance interactions for all tested networks.
%The nodes with high balance tendency is more likely 
%to be influential spreaders.

\subsection*{S1 Text}
\label{S1_Text}
{\bf $k$-shell index.}
%In order to assign $k$-shell index, we remove all nodes with degree less than $k$
%until no nodes less than $k$ remains.
%Then the maximum subgraph whose remained degree is larger than $k$ is called $k$-core.
%$k$-shell is the set of all nodes belonging to the $k$-core but not to the $(k+1)$-core.
%As a result, each node is assigned with a unique $k$-shell index.

\subsection*{S2 Text}
\label{S2_Text}
{\bf Identifying structural hole in the link community.}
\section*{Acknowledgments}
This work was funded by NIH-NIGMS 1R21GM107641.
FL was supported by Riksbankens Jubileumsfond (The Bank of Sweden
Tercentenary Foundation) Grant BR. P12-0705:1.

\nolinenumbers

% Either type in your references using
% \begin{thebibliography}{}
% \bibitem{}
% Text
% \end{thebibliography}
%
% OR
%
% Compile your BiBTeX database using our plos2009.bst
% style file and paste the contents of your .bbl file
% here.

%\section*{References}

\clearpage

\section*{Supporting information}
\subsection*{Identifying structural hole}
In order to identify the intercommunity links, called the structural hole, 
we use the link community detection method proposed in \cite{ahn}.
We adapt the method for local version only using local information 
of networks because global information is difficult to obtain by surveys.
The method for identifying structural hole link is following.
When a new link $e_{ik}$ is added, 
likewise the original link community algorithm \cite{ahn}, 
we define the similarity $S(e_{ik},e_{jk})$ between two links 
$e_{ik}$ and each of existed links $e_{jk}$ by following,
\begin{equation}
S(e_{ik},e_{jk})=\frac{|n_+(i) \cap n_+(j)|}{ |n_+(i) \cup n_+(j)|},
\end{equation}
where $n_+(i)$ is the set of neighbors of node $i$.
Therefore, if there are many common friends 
the similarity is high.
Then, if the similarity is less than a certain threshold 
meaning that two neighbors have only few fraction of common friends,
we judge the newly added link as a structural hole.

\subsection*{Epidemic size for a seed node with degree $k$ on randomized networks}
The susceptible-infected-recovered (SIR) model on a network can be mapped into bond percolation problem with the probability
of link occupation $\beta$.
In the perspective of bond percolation, the epidemic size initiated by a single seed node is the statistically same as the average 
size of component including the seed node.
Then, one can obtain the epidemic size initiated by a seed node having degree $k$ 
by following a generating function method \cite{newman_gf,newman_epi}.
Given the degree distribution $q(k)$ of a network, we define the degree generating function as 
$G_0(x)=\sum_{k=0}^{\infty} q(k)x^k$.
We also define the generating function for the excess degree, for a node reached by following a randomly 
chosen link, as $G_1(x)=\frac{1}{\langle k\rangle} \frac{d G_0(x)}{dx}$.
For the locally-tree like networks, the probability $u$ that a node reached by following a randomly chosen link
does not belong to the giant component is given by
\begin{equation}
u=\sum_{k=1}^{\infty} \frac{k q(k)}{\langle k\rangle}[1+(u-1)\beta]^{k-1}=G_1[1+(u-1)\beta].
\end{equation}
The probability $p_k$ that a randomly chosen node with degree $k$ belongs to the giant component can be obtained as
\begin{equation}
p_k=1-[1+(u-1)\beta]^k.
\end{equation}
We can also obtain the size $s$ of the giant component of a given network as $s=1-G_0[1+(u-1)\beta]$.
Finally, the average epidemic size, $S_k$, initiated by node with a degree $k$ is a 
product of $p_k$ and $s$, $S_k=s p_k$.

\section*{Tables}
\begin{landscape}
\begin{table}[t]
\caption{
	{\bf Multilinear regression for the QXF network.} 
$M_i=c_0+c_1 a^{exc}_i +c_2 a^{bal}_i +c_3 a^{ca}_i + c_4 a^{sh}_i + c_5 k_i +c_6 k^{sh}_i +c_7 k^{2sum}_i +c_8 k^{sum}_i+\epsilon.$
}
\begin{tabular}{|l|l|l|l|l|l|l|l|l|l|}
\hline
intercept & $k_{sh}$ & $k_{2sum}$ & $k_{sum}$ & $k$ & exchange & balance & col. act. &  str. hole & $R^2$\\ 
\hline
$-0.953^{***}$ & $0.118^{***}$ & $6.82e$-6$^{***}$ & $5.63e$-5$^{***}$ & $4.41e$-3$^{***}$ & $-0.353^{***}$ & $0.330^{***}$ & $0.153^{*}$ & $-0.379^{***}$ & 0.950 \\ % all
\hdashline
$[0.0199]$ & $[1.41e$-3$]$ & $[1.22e$-7$]$ & $[2.76e$-6$]$ & $[1.65e$-4$]$ & $[0.0253]$ & $[0.0258]$ & $[0.0390]$ & $[0.0781]$ & -  \\ 
\hline
$-0.0859^{***}$ &- & $1.49e$-5$^{***}$ & $-2.51e$-5$^{***}$ & $6.01e$-3$^{***}$ & $-0.374^{***}$ & $0.766^{***}$ & $0.0890^{NS}$ & $-1.02^{***}$ & 0.923 \\ % ksh
\hdashline
$[0.0211]$ &- & $[9.22e$-8$]$ & $[3.19e$-6$]$ & $[2.02e$-4$]$ & $[0.0313]$ & $[0.0312]$ & $[0.0482]$ & $[0.0962]$ & - \\ 
\hline
$-1.20^{***}$ & $0.181^{***}$ &- & $1.64e$-4$^{***}$ & $2.40e$-3$^{***}$ & $-0.367^{***}$ & $0.103^{***}$ & $0.359^{***}$ & $-0.298^{***}$ & 0.938 \\ % k2sum
\hdashline
$[0.0216]$ & $[9.64e$-4$]$ &- & $[2.21e$-6$]$ & $[1.79e$-4$]$ & $[0.0281]$ & $[0.0283]$ & $[0.0432]$ & $[0.0869]$ & - \\ 
\hline
$-0.983^{***}$ & $0.108^{***}$ & $8.54e$-6$^{***}$ &- & $6.88e$-3$^{***}$ & $-0.338^{***}$ & $0.388^{***}$ & $0.198^{***}$ & $-0.414^{***}$ & 0.948 \\ % ksum
\hdashline
$[0.0202]$ & $[1.34e$-3$]$ & $[8.89e$-8$]$ &- & $[1.14e$-4$]$ & $[0.0256]$ & $[0.0260]$ & $[0.0395]$ & $[0.0793]$ & - \\ 
\hline
$-0.868^{***}$ & $0.123^{***}$ & $6.11e$-6$^{***}$ & $1.11e$-4$^{***}$ &- & $-0.402^{***}$ & $0.350^{***}$ & $-0.0573^{NS}$ & $-0.330^{***}$ & 0.947 \\ % k 
\hdashline
$[0.0202]$ & $[1.44e$-3$]$ & $[1.22e$-7$]$ & $[1.92e$-6$]$ &- & $[0.0259]$ & $[0.0264]$ & $[0.0392]$ & $[0.0801]$ & - \\ 
\hline
$-1.03^{***}$ & $0.118^{***}$ & $6.84e$-6$^{***}$ & $5.52e$-5$^{***}$ & $4.58e$-3$^{***}$ &- & $0.264^{***}$ & $0.197^{***}$ & $-0.426^{***}$ & 0.949 \\ % exchange
\hdashline
$[0.0194]$ & $[1.42e$-3$]$ & $[1.22e$-7$]$ & $[2.78e$-6$]$ & $[1.65e$-4$]$ &- & $[0.0255]$ & $[0.0391]$ & $[0.0786]$ & - \\ 
\hline
$-0.961^{***}$ & $0.122^{***}$ & $6.58e$-6$^{***}$ & $6.02e$-5$^{***}$ & $4.47e$-3$^{***}$ & $-0.294^{***}$ &- & $0.235^{***}$ & $-0.511^{***}$ & 0.949 \\ % balance
\hdashline
$[0.0200]$ & $[1.39e$-3$]$ & $[1.21e$-7$]$ & $[2.76e$-6$]$ & $[1.66e$-4$]$ & $[0.0250]$ &- & $[0.0387]$ & $[0.0779]$ & - \\ 
\hline
$-0.907^{***}$ & $0.118^{***}$ & $6.87e$-6$^{***}$ & $5.70e$-5$^{***}$ & $4.28e$-3$^{***}$ & $-0.361^{***}$ & $0.347^{***}$ &- & $-0.313^{***}$ & 0.950 \\ % col
\hdashline
$[0.0161]$ & $[1.41e$-3$]$ & $[1.21e$-7$]$ & $[2.76e$-6$]$ & $[1.61e$-4$]$ & $[0.0252]$ & $[0.0254]$ &- & $[0.0763]$ & - \\ 
\hline
$-0.963^{***}$ & $0.119^{***}$ & $6.81e$-6$^{***}$ & $5.66e$-5$^{***}$ & $4.39e$-3$^{***}$ & $-0.358^{***}$ & $0.347^{***}$ & $-0.112^{**}$ &- & 0.950 \\ % hole
\hdashline
$[0.0199]$ & $[1.41e$-3$]$ & $[1.22e$-7$]$ & $[2.76e$-6$]$ & $[1.65e$-4$]$ & $[0.0253]$ & $[0.0256]$ & $[0.0381]$ &- & - \\ 
\hline
$-2.05^{***}$ & $0.281^{***}$ & - &- &- &- &- &- &- & 0.823 \\ % ksh
\hdashline
$[0.0247]$ & $[0.00114]$ &- &- &- &- &- &- &- & - \\ 
\hline
$-0.121^{***}$ &- & $1.64e$-5$^{***}$ &- &- &- &- &- &- & 0.908 \\ % k2sum
\hdashline
$[0.0126]$ &- & $[4.54$-8$]$ &- &- &- &- &- &- &- \\ 
\hline
$1.49^{***}$ &- &- & $3.54e$-4$^{***}$ &- &- &- &- &- & 0.757 \\ % ksum
\hdashline
$[0.0160]$ &- &- & $[1.75$-6$]$ &- &- &- &- &- & -\\ 
\hline
$1.92^{***}$ &- &- &- & $0.0280^{***}$ &- &- &- &- & 0.580 \\ % k
\hdashline
$[0.0202]$ &- &- &- & $[0.000208]$ &- &- &- &- & \\ 
\hline
\end{tabular}
\begin{flushleft} 
	Coefficients [standard errors] for the linear models. 
	Here, $***$ denotes $p<0.001$, $**$ denotes $p<0.01$, $*$ denotes $p<0.05$, and $NS$ denotes not significant with significant level of p-value $0.05$.
\end{flushleft}
\label{tab:table2}
\end{table}

%%%%% Guestbook
\begin{table}[!ht]
\caption{
	\bf{Multilinear regression for the QXG network.} $M_i=c_0+c_1 a^{exc}_i +c_2 a^{bal}_i +c_3 a^{ca}_i + c_4 a^{sh}_i + c_5 k_i +c_6 k^{sh}_i +c_7 k^{2sum}_i +c_8 k^{sum}_i+\epsilon.$
}
\begin{tabular}{|c|c|c|c|c|c|c|c|c|c|}
\hline
intercept & $k_{sh}$ & $k_{2sum}$ & $k_{sum}$ & $k$ & exchange & balance & coll. act. &  str. hole & $R^2$\\ 
\hline
$-0.831^{***}$ & $0.238^{***}$ & $5.97e$-5$^{***}$ & $-5.16e$-4$^{***}$ & $1.20e$-2$^{***}$ & $-4.02e$-3$^{NS}$ & $0.400^{***}$ & $0.157^{***}$ & $-0.295^{***}$ & 0.966 \\ % all
\hdashline
$[0.0206]$ & $[2.67e$-3$]$ & $[7.71e$-7$]$ & $[1.88e$-5$]$ & $[3.86e$-4$]$ & $[0.0210]$ & $[0.0276]$ & $[0.0435]$ & $[0.0580]$ & \\ 
\hline
$0.268^{***}$ & & $1.09e$-4$^{***}$ & $-1.38e$-3$^{***}$ & $1.79e$-2$^{***}$ & $0.248^{***}$ & $0.951^{***}$ & $0.339^{***}$ & $-0.852^{***}$ & 0.934 \\ % ksh
\hdashline
$[0.0228]$ & & $[7.45e$-7$]$ & $[2.22e$-5$]$ & $[5.26e$-4$]$ & $[0.0288]$ & $[0.0372]$ & $[0.0601]$ & $[0.0797]$ & \\ 
\hline
$-1.31^{***}$ & $0.386^{***}$ & & $7.48e$-4$^{***}$ & $5.66e$-3$^{***}$ & $-0.179^{***}$ & $0.0961^{**}$ & $0.436^{***}$ & $-2.27^{**}$ & 0.942 \\ % k2sum
\hdashline
$[0.0254]$ & $[2.42e$-3$]$ & & $[1.20e$-5$]$ & $[4.90e$-4$]$ & $[0.0271]$ & $[0.0354]$ & $[0.0563]$ & $[0.0753]$ & \\ 
\hline
$-0.902^{***}$ & $0.277^{***}$ & $4.12e$-5$^{***}$ & & $5.67e$-3$^{***}$ & $-0.0801^{***}$ & $0.313^{***}$ & $0.149^{**}$ & $-2.63^{***}$ & 0.963 \\ % ksum
\hdashline
$[0.0213]$ & $[2.38e$-3$]$ & $[3.95e$-7$]$ & & $[3.24e$-4$]$ & $[0.0217]$ & $[0.0285]$ & $[0.0453]$ & $[0.0604]$ & \\ 
\hline
$-0.764^{***}$ & $0.253^{***}$ & $5.46e$-5$^{***}$ & $-1.72e$-4$^{***}$ & & $-0.104^{***}$ & $0.392^{***}$ & $-0.0513^{NS}$ & $-0.270^{***}$ & 0.962 \\ % k 
\hdashline
$[0.0215]$ & $[2.77e$-3$]$ & $[7.94e$-7$]$ & $[1.59e$-5$]$ & & $[0.0219]$ & $[0.0290]$ & $[0.0453]$ & $[0.0611]$ & \\ 
\hline
$-0.832^{***}$ & $0.238^{***}$ & $5.97e$-5$^{***}$ & $-5.17e$-4$^{***}$ & $1.20e$-2$^{***}$ & & $0.400^{***}$ & $0.155^{***}$ & $-0.295^{***}$ & 0.966 \\ % exchange
\hdashline
$[0.0205]$ & $[2.65e$-3$]$ & $[7.67e$-6$]$ & $[1.86e$-5$]$ & $[3.81e$-4$]$ & & $[0.0274]$ & $[0.0423]$ & $[0.0579]$ & \\ 
\hline
$-0.864^{***}$ & $0.247^{***}$ & $5.81e$-5$^{***}$ & $-4.85e$-4$^{***}$ & $1.19e$-2$^{***}$ & $0.315^{NS}$ & & $0.285^{***}$ & $-0.449^{***}$ & 0.965 \\ % balance
\hdashline
$[0.0207]$ & $[2.63e$-3$]$ & $[7.72e$-7$]$ & $[1.89e$-5$]$ & $[3.91e$-4$]$ & $[0.0211]$ & & $[0.0431]$ & $[0.0577]$ & \\ 
\hline
$-0.796^{***}$ & $0.239^{***}$ & $5.99e$-5$^{***}$ & $-5.16e$-4$^{***}$ & $1.17e$-2$^{***}$ & $0.0137^{NS}$ & $0.420^{***}$ & & $-2.20^{***}$ & 0.966 \\ % col
\hdashline
$[0.0182]$ & $[2.67e$-3$]$ & $[7.69e$-7$]$ & $[1.88e$-5$]$ & $[3.82e$-4$]$ & $[0.0204]$ & $[0.0270]$ & & $[0.0542]$ & \\ 
\hline
$-0.833^{***}$ & $0.240^{***}$ & $5.96e$-5$^{***}$ & $-5.14e$-4$^{***}$ & $1.19e$-2$^{***}$ & $-0.00865^{NS}$ & $0.426^{***}$ & $0.0781^{NS}$ & & 0.966 \\ % hole
\hdashline
$[0.0206]$ & $[2.66e$-3$]$ & $[7.72e$-7$]$ & $[1.88e$-5$]$ & $[3.86e$-4$]$ & $[0.0210]$ & $[0.0271]$ & $[0.0407]$ & & \\ 
\hline
$-2.22^{***}$ & $0.600^{***}$ & & & & & & & & 0.815 \\ % ksh
\hdashline
$[0.0351]$ & $[0.00306]$ & & & & & & & & \\ 
\hline
$0.918^{***}$ & & $7.08e$-5$^{***}$ & & & & & & & 0.896 \\ % k2sum
\hdashline
$[0.0160]$ & & $[2.58$-7$]$ & & & & & & & \\ 
\hline
$1.79^{***}$ & & & $1.58e$-3$^{***}$ & & & & & & 0.750 \\ % ksum
\hdashline
$[0.0224]$ & & & $[9.76$-6$]$ & & & & & & \\ 
\hline
$2.09^{***}$ & & & & $0.0628^{***}$ & & & & & 0.624 \\ % k
\hdashline
$[0.0269]$ & & & & $[0.000522]$ & & & & & \\ 
\hline
\end{tabular}
\begin{flushleft} 
	Coefficients [standard errors] for the linear models. 
	Every row corresponds to a different model. Here, $***$ denotes $p<0.001$, $**$ denotes $p<0.01$, $*$ denotes $p<0.05$, and $NS$ denotes not significant.
\end{flushleft}
\label{tab:guest1}
\end{table}

%%%%% POK Message
\begin{table}[!ht]
\caption{
	\bf{Multilinear regression for the POK network.} $M_i=c_0+c_1 a^{exc}_i +c_2 a^{bal}_i +c_3 a^{ca}_i + c_4 a^{sh}_i + c_5 k_i +c_6 k^{sh}_i +c_7 k^{2sum}_i +c_8 k^{sum}_i.$
}
\begin{tabular}{|c|c|c|c|c|c|c|c|c|c|}
\hline
intercept & $k_{sh}$ & $k_{2sum}$ & $k_{sum}$ & $k$ & exchange & balance & coll. act. &  str. hole & $R^2$\\ 
\hline
$-1.19^{***}$ & $0.457^{***}$ & $2.62e$-5$^{***}$ & $-1.99e$-4$^{***}$ & $7.88e$-3$^{***}$ & $-0.0827^{*}$ & $-0.0819^{*}$ & $0.813^{***}$ & $-0.594^{***}$ & 0.907 \\ % all
\hdashline
$[0.0159]$ & $[2.56e$-3$]$ & $[1.03e$-6$]$ & $[1.14e$-5$]$ & $[1.71e$-4$]$ & $[0.0205]$ & $[0.0401]$ & $[0.0360]$ & $[0.0456]$ & \\ 
\hline
$-0.679^{***}$ & & $1.63e$-4$^{***}$ & $-1.50e$-3$^{***}$ & $2.33e$-2$^{***}$ & $0.165^{***}$ & $0.651^{***}$ & $1.05^{***}$ & $-1.72^{***}$ & 0.727 \\ % ksh
\hdashline
$[0.0267]$ & & $[1.19e$-6$]$ & $[1.51e$-5$]$ & $[2.52e$-4$]$ & $[0.0350]$ & $[0.0682]$ & $[0.0614]$ & $[0.0773]$ & \\ 
\hline
$-1.16^{***}$ & $0.505^{***}$ & & $8.02e$-5$^{***}$ & $5.51e$-3$^{***}$ & $-0.0873^{***}$ & $-0.106^{**}$ & $0.887^{***}$ & $-0.509^{***}$ & 0.903 \\ % k2sum
\hdashline
$[0.0161]$ & $[1.75e$-3$]$ & & $[3.11e$-6$]$ & $[1.45e$-4$]$ & $[0.0209]$ & $[0.0408]$ & $[0.0365]$ & $[0.0464]$ & \\ 
\hline
$-1.16^{***}$ & $0.485^{***}$ & $8.83e$-6$^{***}$ & & $6.12e$-3$^{***}$ & $-0.0881^{***}$ & $-0.111^{**}$ & $0.853^{***}$ & $-5.70^{***}$ & 0.905 \\ % ksum
\hdashline
$[0.0159]$ & $[1.99e$-3$]$ & $[2.78e$-7$]$ & & $[1.39e$-4$]$ & $[0.0207]$ & $[0.0404]$ & $[0.0362]$ & $[0.0460]$ & \\ 
\hline
$-1.16^{***}$ & $0.516^{***}$ & $4.95e$-8$^{NS}$ & $1.12e$-4$^{***}$ & & $-0.107^{***}$ & $-0.110^{**}$ & $0.777^{***}$ & $-0.609^{***}$ & 0.895 \\ % k 
\hdashline
$[0.0168]$ & $[2.34e$-3$]$ & $[9.17e$-7$]$ & $[9.81e$-6$]$ & & $[0.0218]$ & $[0.0426]$ & $[0.0382]$ & $[0.0485]$ & \\ 
\hline
$-1.21^{***}$ & $0.456^{***}$ & $2.62e$-5$^{***}$ & $-2.00e$-4$^{***}$ & $7.89e$-3$^{***}$ & & $-0.0952^{*}$ & $0.807^{***}$ & $-0.599^{***}$ & 0.906 \\ % exchange
\hdashline
$[0.0151]$ & $[2.55e$-3$]$ & $[1.03e$-6$]$ & $[1.14e$-5$]$ & $[1.71e$-4$]$ & & $[0.0400]$ & $[0.0360]$ & $[0.0456]$ & \\ 
\hline
$-1.19^{***}$ & $0.456^{***}$ & $2.62e$-5$^{***}$ & $-2.00e$-4$^{***}$ & $7.88e$-3$^{***}$ & $-0.0861^{***}$ & & $0.809^{***}$ & $-0.586^{***}$ & 0.907 \\ % balance
\hdashline
$[0.0159]$ & $[2.54e$-3$]$ & $[1.03e$-6$]$ & $[1.14e$-5$]$ & $[1.71e$-4$]$ & $[0.0204]$ & & $[0.0359]$ & $[0.0455]$ & \\ 
\hline
$0.963^{***}$ & $0.459^{***}$ & $2.81e$-5$^{***}$ & $-2.16e$-4$^{***}$ & $7.79e$-3$^{***}$ & $-0.0620^{**}$ & $-0.0322^{NS}$ & & $-0.229^{***}$ & 0.904 \\ % col
\hdashline
$[0.0126]$ & $[2.59e$-3$]$ & $[1.04e$-6$]$ & $[1.16e$-5$]$ & $[1.73e$-4$]$ & $[0.0208]$ & $[0.0406]$ & & $[0.0433]$ & \\ 
\hline
$-1.17^{***}$ & $0.461^{***}$ & $2.52e$-5$^{***}$ & $-1.95e$-4$^{***}$ & $7.89e$-3$^{***}$ & $-0.0906^{***}$ & $0.0399^{NS}$ & $0.648^{***}$ & & 0.906 \\ % hole
\hdashline
$[0.0159]$ & $[2.54e$-3$]$ & $[1.03e$-6$]$ & $[1.15e$-5$]$ & $[1.72e$-4$]$ & $[0.0206]$ & $[0.0402]$ & $[0.0338]$ & & \\ 
\hline
$-1.00^{***}$ & $0.555^{***}$ & & & & & & & & 0.883 \\ % ksh
\hdashline
$[0.0115]$ & $[0.00157]$ & & & & & & & & \\ 
\hline
$0.638^{***}$ & & $5.81e$-5$^{***}$ & & & & & & & 0.481 \\ % k2sum
\hdashline
$[0.0196]$ & & $[4.69$-7$]$ & & & & & & & \\ 
\hline
$1.42^{***}$ & & & $5.91e$-4$^{***}$ & & & & & & 0.326 \\ % ksum
\hdashline
$[0.0189]$ & & & $[6.59$-6$]$ & & & & & & \\ 
\hline
$2.14^{***}$ & & & & $0.0230^{***}$ & & & & & 0.180 \\ % k
\hdashline
$[0.0173]$ & & & & $[0.000380]$ & & & & & \\ 
\hline
\end{tabular}
\begin{flushleft} 
	Coefficients [standard errors] for the linear models. 
	Every row corresponds to a different model. Here, $***$ denotes $p<0.001$, $**$ denotes $p<0.01$, $*$ denotes $p<0.05$, and $NS$ denotes not significant.
\end{flushleft}
\label{tab:pok1}
\end{table}

%%%%% Livejournal
\begin{table}[!ht]
\caption{
	\bf{Multilinear regression for the LJ network.} $M_i=c_0+c_1 a^{exc}_i +c_2 a^{bal}_i +c_3 a^{ca}_i + c_4 a^{sh}_i + c_5 k_i +c_6 k^{sh}_i +c_7 k^{2sum}_i +c_8 k^{sum}_i.$
}
\begin{tabular}{|c|c|c|c|c|c|c|c|c|c|}
\hline
intercept & $k_{sh}$ & $k_{2sum}$ & $k_{sum}$ & $k$ & exchange & balance & coll. act. &  str. hole & $R^2$\\ 
\hline
$0.0698^{***}$ & $0.0887^{***}$ & $7.79e$-6$^{***}$ & $-5.76e$-5$^{***}$ & $1.61e$-3$^{***}$ & $0.386^{*}$ & $0.307^{***}$ & $-2.39^{***}$ & $-0.129^{***}$ & 0.682 \\ % all
\hdashline
$[0.0138]$ & $[5.41e$-4$]$ & $[4.34e$-8$]$ & $[5.78e$-7$]$ & $[3.93e$-5$]$ & $[0.0699]$ & $[0.0366]$ & $[0.0345]$ & $[0.0373]$ & \\ 
\hline
$0.396^{***}$ & & $7.99e$-6$^{***}$ & $-4.51e$-5$^{***}$ & $2.48e$-3$^{***}$ & $0.912^{***}$ & $2.99^{***}$ & $-0.277^{***}$ & $-2.11^{***}$ & 0.551 \\ % ksh
\hdashline
$[0.0162]$ & & $[5.16e$-8$]$ & $[6.81e$-7$]$ & $[4.63e$-5$]$ & $[0.0830]$ & $[0.0389]$ & $[0.0411]$ & $[0.0439]$ & \\ 
\hline
$0.782^{***}$ & $0.0913^{***}$ & & $3.43e$-5$^{***}$ & $-1.76e$-5$^{NS}$ & $0.301^{***}$ & $0.721^{***}$ & $-0.0732^{NS}$ & $0.343^{***}$ & 0.525 \\ % k2sum
\hdashline
$[0.0161]$ & $[6.61e$-4$]$ & & $[3.27e$-7$]$ & $[4.67e$-5$]$ & $[0.0855]$ & $[0.0446]$ & $[0.0422]$ & $[0.0442]$ & \\ 
\hline
$0.431^{***}$ & $0.0816^{***}$ & $3.95e$-6$^{***}$ & & $3.99e$-4$^{***}$ & $0.292^{***}$ & $0.552^{***}$ & $-0.361^{***}$ & $-0.981^{***}$ & 0.634 \\ % ksum
\hdashline
$[0.0143]$ & $[5.76e$-4$]$ & $[2.16e$-8$]$ & & $[4.01e$-5$]$ & $[0.0750]$ & $[0.0392]$ & $[0.0371]$ & $[0.0399]$ & \\ 
\hline
$0.128^{***}$ & $0.0916^{***}$ & $7.38e$-6$^{***}$ & $-5.03e$-5$^{***}$ & & $0.344^{***}$ & $0.332^{***}$ & $-0.342^{***}$ & $-1.30^{***}$ & 0.674 \\ % k 
\hdashline
$[0.0139]$ & $[5.43e$-4$]$ & $[4.28e$-8$]$ & $[5.56e$-7$]$ & & $[0.0708]$ & $[0.0370]$ & $[0.0349]$ & $[0.0377]$ & \\ 
\hline
$0.0768^{***}$ & $0.0888^{***}$ & $7.79e$-6$^{***}$ & $-5.75e$-5$^{***}$ & $1.61e$-3$^{***}$ & & $0.331^{***}$ & $-0.224^{***}$ & $-1.28^{***}$ & 0.681 \\ % exchange
\hdashline
$[0.0137]$ & $[5.40e$-4$]$ & $[4.34e$-8$]$ & $[5.78e$-7$]$ & $[3.93e$-5$]$ & & $[0.0363]$ & $[0.0345]$ & $[0.0373]$ & \\ 
\hline
$0.0558^{***}$ & $0.0907^{***}$ & $7.81e$-6$^{***}$ & $-5.79e$-5$^{***}$ & $1.62e$-3$^{***}$ & $0.455^{***}$ & & $-0.152^{***}$ & $-1.33^{**}$ & 0.682 \\ % balance
\hdashline
$[0.0137]$ & $[4.84e$-4$]$ & $[4.34e$-8$]$ & $[5.77e$-7$]$ & $[3.93e$-5$]$ & $[0.0694]$ & & $[0.0329]$ & $[0.0370]$ & \\ 
\hline
$0.00122^{NS}$ & $0.0887^{***}$ & $7.78e$-6$^{***}$ & $-5.77e$-5$^{***}$ & $1.63e$-4$^{***}$ & $0.400^{***}$ & $0.231^{***}$ & & $-0.136^{*}$ & 0.682 \\ % col
\hdashline
$[0.00956]$ & $[5.41e$-4$]$ & $[4.34e$-8$]$ & $[5.77e$-7$]$ & $[3.92e$-5$]$ & $[0.0699]$ & $[0.0349]$ & & $[0.0357]$ & \\ 
\hline
$-0.103^{***}$ & $0.0912^{***}$ & $7.43e$-6$^{***}$ & $-5.59e$-5$^{***}$ & $1.62e$-3$^{***}$ & $0.331^{***}$ & $0.475^{***}$ & $-0.584^{***}$ & & 0.676 \\ % hole
\hdashline
$[0.0139]$ & $[5.41e$-4$]$ & $[4.25e$-8$]$ & $[5.81e$-7$]$ & $[3.96e$-5$]$ & $[0.0705]$ & $[0.0366]$ & $[0.0334]$ & & \\ 
\hline
%%%%%%%%%%%%%%%%%%%%%%%%%%%%%%%%%%%%%%%%%%%%%%%%
$1.17^{***}$ & $0.123^{***}$ & & & & & & & & 0.420 \\ % ksh
\hdashline
$[0.00986]$ & $[0.000568]$ & & & & & & & & \\ 
\hline
$0.713^{***}$ & & $4.99e$-6$^{***}$ & & & & & & & 0.436 \\ % k2sum
\hdashline
$[0.0109]$ & & $[2.22e$-8$]$ & & & & & & & \\ 
\hline
$1.40^{***}$ & & & $5.57e$-5$^{***}$ & & & & & & 0.310 \\ % ksum
\hdashline
$[0.0105]$ & & & $[3.26e$-7$]$ & & & & & & \\ 
\hline
$2.27^{***}$ & & & & $0.00363^{***}$ & & & & & 0.0501 \\ % k
\hdashline
$[0.0106]$ & & & & $[6.20e$-5$]$ & & & & & \\ 
\hline
\end{tabular}
\begin{flushleft} 
\end{flushleft}
\label{tab:lj1}
\end{table}

%%%%% PROSTITUTION
\begin{table}[!ht]
\caption{
	\bf{Multilinear regression for the PRO network.} $M_i=c_0+c_1 a^{exc}_i +c_2 a^{bal}_i +c_3 a^{ca}_i + c_4 a^{sh}_i + c_5 k_i +c_6 k^{sh}_i +c_7 k^{2sum}_i +c_8 k^{sum}_i.$
}
\begin{tabular}{|c|c|c|c|c|c|c|c|}
\hline
intercept & $k_{sh}$ & $k_{2sum}$ & $k_{sum}$ & $k$ & coll. act. &  str. hole & $R^2$\\ 
\hline
$-1.12^{***}$ & $0.610^{***}$ & $3.76e$-4$^{***}$ & $1.24e$-4$^{***}$ & $1.25e$-2$^{***}$ & $1.13^{***}$ & $-0.630^{***}$ & 0.945 \\ % all
\hdashline
$[0.0204]$ & $[5.36e$-3$]$ & $[5.09e$-6$]$ & $[3.58e$-5$]$ & $[7.59e$-4$]$ & $[0.0361]$ & $[0.0556]$ & \\ 
\hline
$0.371^{***}$ & & $6.35e$-4$^{***}$ & $9.69e$-4$^{***}$ & $2.74e$-2$^{***}$ & $0.604^{***}$ & $0.626^{***}$ & 0.866 \\ % ksh 
\hdashline
$[0.0242]$ & & $[7.08e$-6$]$ & $[5.45e$-5$]$ & $[1.16e$-3$]$ & $[0.0557]$ & $[0.0846]$ & \\ 
\hline
$-1.58^{***}$ & $0.787^{***}$ & & $1.29e$-3$^{***}$ & $2.75e$-2$^{***}$ & $1.69^{***}$ & $0.107^{NS}$ & 0.912 \\ % k2sum
\hdashline
$[0.0246]$ & $[6.06e$-3$]$ & & $[4.06e$-5$]$ & $[9.24e$-4$]$ & $[0.0446]$ & $[0.0691]$ & \\ 
\hline
$-1.15^{***}$ & $0.614^{***}$ & $3.84e$-4$^{***}$ & & $1.32e$-2$^{***}$ & $1.164^{***}$ & $-0.659^{***}$ & 0.945 \\ % ksum
\hdashline
$[0.0193]$ & $[5.25e$-3$]$ & $[4.57e$-6$]$ & & $[7.40e$-4$]$ & $[0.0347]$ & $[0.0550]$ & \\ 
\hline
$-1.05^{***}$ & $0.625^{***}$ & $3.99e$-4$^{***}$ & $2.58e$-4$^{***}$ & & $0.899^{***}$ & $-0.789^{***}$ & 0.943 \\ % k
\hdashline
$[0.0203]$ & $[5.36e$-3$]$ & $[4.98e$-6$]$ & $[3.54e$-5$]$ & & $[0.0338]$ & $[0.0555]$ & \\ 
\hline
$-0.672^{***}$ & $0.589^{***}$ & $4.09e$-4$^{***}$ & $4.38e$-4$^{**}$ & $3.45e$-3$^{***}$ & & $-0.316^{***}$ & 0.939 \\ % a4
\hdashline
$[0.0152]$ & $[5.59e$-3$]$ & $[5.24e$-6$]$ & $[3.62e$-5$]$ & $[7.37e$-4$]$ & & $[0.0575]$ & \\ 
\hline
$-1.10^{***}$ & $0.598^{***}$ & $3.66e$-4$^{***}$ & $1.85e$-4$^{***}$ & $1.40e$-2$^{***}$ & $1.05^{***}$ & & 0.944 \\ % a5
\hdashline
$[0.0204]$ & $[5.29e$-3$]$ & $[5.04e$-6$]$ & $[3.56e$-5$]$ & $[7.53e$-4$]$ & $[0.0358]$ & & \\ 
\hline
$-1.44^{***}$ & $1.07^{***}$ & & & & & & 0.859 \\ % ksh
\hdashline
$[0.0199]$ & $[0.00456]$ & & & & & & \\ 
\hline
$0.621^{***}$ & & $8.40e$-4$^{***}$ & & & & & 0.848 \\ % k2sum
\hdashline
$[0.0138]$ & & $[3.71e$-6$]$ & & & & & \\ 
\hline
$1.25^{***}$ & & & $6.12e$-3$^{***}$ & & & & 0.669 \\ % ksum
\hdashline
$[0.0186]$ & & & $[4.50e$-5$]$ & & & & \\ 
\hline
$1.52^{***}$ & & & & $0.134^{***}$ & & & 0.452 \\ % k
\hdashline
$[0.0238]$ & & & & $[0.00154]$ & & & \\ 
\hline
\end{tabular}
\begin{flushleft} %$\beta=0.04$.
\end{flushleft}
\label{tab:pro1}
\end{table}

\end{landscape}

%%%%%%%%%%%%%%%%%%%%%%% k_i & social mechanisms 
%\begin{table}[t]
%\begin{adjustwidth}{-2.25in}{0in} % Comment out/remove adjustwidth environment if table fits in text column.
%\caption{
%	{\bf Multilinear regression for the QXF network.} 
%$M_i=c_0+c_1 a^{exc}_i +c_2 a^{bal}_i +c_3 a^{ca}_i + c_4 a^{sh}_i + c_5 k_i +c_6 k^{sh}_i +c_7 k^{2sum}_i +c_8 k^{sum}_i+\epsilon.$
%}
%\begin{tabular}{|l|l|l|l|l|l|l|l|l|l|}
%\hline
%intercept & $k_{sh}$ & $k_{2sum}$ & $k_{sum}$ & $k$ & exchange & balance & col. act. &  str. hole & $R^2$\\ 
%\hline
%$-2.05^{***}$ & $0.281^{***}$ & - &- &- &- &- &- &- & 0.823 \\ % ksh
%\hdashline
%$[0.0247]$ & $[0.00114]$ &- &- &- &- &- &- &- & - \\ 
%\hline
%$-0.121^{***}$ &- & $1.64e$-5$^{***}$ &- &- &- &- &- &- & 0.908 \\ % k2sum
%\hdashline
%$[0.0126]$ &- & $[4.54$-8$]$ &- &- &- &- &- &- &- \\ 
%\hline
%$1.49^{***}$ &- &- & $3.54e$-4$^{***}$ &- &- &- &- &- & 0.757 \\ % ksum
%\hdashline
%$[0.0160]$ &- &- & $[1.75$-6$]$ &- &- &- &- &- & -\\ 
%\hline
%$1.92^{***}$ &- &- &- & $0.0280^{***}$ &- &- &- &- & 0.580 \\ % k
%\hdashline
%$[0.0202]$ &- &- &- & $[0.000208]$ &- &- &- &- & \\ 
%\hline
%\end{tabular}
%\begin{flushleft} 
%	Coefficients [standard errors] for the linear models. 
%	Here, $***$ denotes $p<0.001$, $**$ denotes $p<0.01$, $*$ denotes $p<0.05$, and $NS$ denotes not significant with significant level of $0.05$.
%\end{flushleft}
%\label{tab:table2}
%\end{adjustwidth}
%\end{table}

\begin{table}[!ht]
\caption{
	\bf{Multilinear regression for the QXF network.} $M_i=c_0+c_1 a^{exc}_i +c_2 a^{bal}_i +c_3 a^{ca}_i + c_4 a^{sh}_i + c_5 k_i.$
}
\begin{tabular}{|c|c|c|c|c|c|c|}
\hline
intercept & $k$ & exchange & balance & coll. act. &  str. hole & $R^2$\\ 
\hline
$0.889^{***}$ & $0.0268^{***}$ & $-0.416^{***}$ & $1.11^{***}$ & $2.89^{***}$ & $-3.79^{***}$ & 0.625 \\ % all
\hdashline
$[0.0432]$ & $[0.000205]$ & $[0.0690]$ & $[0.0688]$ & $[0.103]$ & $[0.211]$ & \\ 
\hline
$0.808^{***}$ & $0.0270^{***}$ &- & $1.04^{***}$ & $2.94^{***}$ & $-3.84^{***}$ & 0.624 \\ % exchange
\hdashline
$[0.0410]$ & $[0.000204]$ & & $[0.0677]$ & $[0.102]$ & $[0.211]$ & \\ 
\hline
$0.954^{***}$ $0.0276^{***}$ & $-0.209^{**}$ &- & $3.20^{***}$ & $-4.34^{***}$ & 0.617 \\ % balance
\hdashline
$[0.0433]$ & $[0.000201]$ & $[0.0685]$ & & $[0.102]$ & $[0.210]$ & \\ 
\hline
$1.86^{***}$ & $0.0265^{***}$ & $-0.579^{***}$ & $1.48^{***}$ &- & $-2.63^{***}$ & 0.602 \\ % col
\hdashline
$[0.0269]$ & $[0.000211]$ & $[0.0708]$ & $[0.0696]$ & & $[0.213]$ & \\ 
\hline
$0.868^{***}$ & $0.0272^{***}$ & $-0.473^{***}$ & $1.32^{***}$ & $2.53^{***}$ &- & 0.615 \\ % a5
\hdashline
$[0.0436]$ & $[0.000207]$ & $[0.0675]$ & $[0.0687]$ & $[0.102]$ & & \\ 
\hline
$2.40^{***}$ &-& $-1.39^{***}$ & $3.26^{***}$ & $2.06^{***}$ & $-6.12^{***}$ & 0.133 \\ % all 
\hdashline
$[0.0631]$ & & $[0.104]$ & $[0.102]$ & $[0.156]$ & $[0.319]$ & \\ 
\hline
$2.15^{***}$  &-&- & $3.04^{***}$ & $2.22^{***}$ & $-6.36^{***}$ & 0.121 \\ % a1
\hdashline
$[0.0607]$ & & & $[0.101]$ & $[0.156]$ & $[0.321]$ & \\ 
\hline
$2.74^{***}$ &-& $-0.843^{***}$ &- & $2.95^{***}$ & $-8.07^{***}$  & 0.0650 \\ % a2 
\hdashline
$[0.0647]$ & & $[0.107]$ & & $[0.159]$ & $[0.326]$ & \\ 
\hline
$3.07^{***}$ &-& $-1.50^{***}$ & $3.50^{***}$ &- & $-5.28^{***}$ & 0.121 \\ % a4 
\hdashline
$[0.0373]$ & & $[0.105]$ & $[0.101]$ & & $[0.314]$ & \\ 
\hline
$2.39^{***}$ &-& $-1.51^{***}$ & $3.63^{***}$ & $1.46^{***}$ &- & 0.109 \\ % a5 
\hdashline
$[0.0640]$ & & $[0.105]$ & $[0.101]$ & $[0.155]$ & & \\ 
\hline
\end{tabular}
\begin{flushleft}
\end{flushleft}
\label{tab:guest2}
\end{table}

%$1.98^{***}$ $[0.0242]$ & $0.0279^{***}$ $[0.000208]$ & $-0.327^{***}$ $[0.0715]$ &-&-&- & 0.581 \\ % k and a1 
%\hline
%$1.66^{***}$ $[0.0233]$ & $0.0269^{***}$ $[0.000212]$ &-& $1.49^{***}$ $[0.0686]$ &-&- & 0.595 \\ % k and a2 
%\hline
%$0.893^{***}$ $[0.0415]$ & $0.0282^{***}$ $[0.000202]$ &-&-& $2.86^{***}$ $[0.102]$ &- & 0.604 \\ % k and a4 
%\hline
%$2.03^{***}$ $[0.0212]$ & $0.0276^{***}$ $[0.000208]$ &-&-&-& $-3.24^{***}$ $[0.215]$  & 0.588 \\ % k and a5 
%\hline
%\end{tabular}
%\begin{flushleft}
%	Coefficients [standard errors] for the linear models. 
%	Every row corresponds to a different model.
%	Here, $***$ denotes $p<0.001$, $**$ denotes $p<0.01$, $*$ denotes $p<0.05$, and $NS$ denotes not significant with significant level of $0.05$.
%\end{flushleft}
%\label{tab:table3}
%\end{table}

\begin{table}[!ht]
\caption{
	\bf{Multilinear regression for the QXG network.} $M_i=c_0+c_1 a^{exc}_i +c_2 a^{bal}_i +c_3 a^{ca}_i + c_4 a^{sh}_i + c_5 k_i.$
}
\begin{tabular}{|c|c|c|c|c|c|c|}
\hline
intercept & $k$ & exchange & balance & coll. act. &  str. hole & $R^2$\\ 
\hline
$0.63^{***}$ & $0.0600^{***}$ & $0.517^{**}$ & $1.25^{***}$ & $3.24^{***}$ & $-2.41^{***}$ & 0.684 \\ % all
\hdashline
$[0.0482]$ & $[0.000490]$ & $[0.0628]$ & $[0.0815]$ & $[0.127]$ & $[0.174]$ & \\ 
\hline
$0.71^{***}$ & $0.0597^{***}$ & & $1.35^{***}$ & $3.53^{***}$ & $-2.38^{***}$ & 0.682 \\ % exchange
\hdashline
$[0.0474]$ & $[0.000491]$ & & $[0.0808]$ & $[0.122]$ & $[0.174]$ & \\ 
\hline
$0.657^{***}$ & $0.0611^{***}$ & $0.668^{***}$ & & $3.71^{***}$ & $-3.00^{***}$ & 0.675 \\ % balance
\hdashline
$[0.0488]$ & $[0.000491]$ & $[0.0629]$ & & $[0.125]$ & $[0.172]$ & \\ 
\hline
$1.46^{***}$ & $0.0608^{***}$ & $0.960^{***}$ & $1.75^{***}$ & & $-0.893^{***}$ & 0.660 \\ % col
\hdashline
$[0.0369]$ & $[0.000507]$ & $[0.0626]$ & $[0.0820]$ & & $[0.170]$ & \\ 
\hline
$0.681^{***}$ & $0.0608^{***}$ & $0.496^{***}$ & $1.50^{***}$ & $2.64^{***}$ & & 0.677 \\ % hole
\hdashline
$[0.0486]$ & $[0.000493]$ & $[0.0635]$ & $[0.0804]$ & $[0.121]$ & & \\ 
\hline
$2.10^{***}$ & & $-0.102^{***}$ & $2.72^{***}$ & $4.23^{***}$ & $-4.81^{***}$ & 0.141 \\ % k
\hdashline
$[0.0769]$ & & $[0.103]$ & $[0.133]$ & $[0.209]$ & $[0.285]$ & \\ 
\hline
$2.08^{***}$ & & & $2.70^{***}$ & $4.17^{***}$ & $-4.82^{***}$ & 0.141 \\ % k a1
\hdashline
$[0.0756]$ & & & $[0.131]$ & $[0.201]$ & $[0.285]$ & \\ 
\hline
$2.21^{***}$ & & $0.208^{***}$ & & $5.31^{***}$ & $-6.22^{***}$ & 0.100 \\ % k a2
\hdashline
$[0.0785]$ & & $[0.104]$ & & $[0.207]$ & $[0.283]$ & \\ 
\hline
$3.21^{***}$ & & $0.467^{***}$ & $3.40^{***}$ & & $-2.86^{***}$ & 0.101 \\ % k c
\hdashline
$[0.0552]$ & & $[0.102]$ & $[0.132]$ & & $[0.275]$ & \\ 
\hline
$2.23^{***}$ & & $-0.161^{NS}$ & $3.26^{***}$ & $3.04^{***}$ & & 0.113 \\ % k a5
\hdashline
$[0.0778]$ & & $[0.105]$ & $[0.131]$ & $[0.200]$ & & \\ 
\hline
\end{tabular}
\begin{flushleft}
\end{flushleft}
\label{tab:guest2}
\end{table}

\begin{table}[!ht]
\caption{
	\bf{Multilinear regression for the POK network.} $M_i=c_0+c_1 a^{exc}_i +c_2 a^{bal}_i +c_3 a^{ca}_i + c_4 a^{sh}_i + c_5 k_i.$
}
\begin{tabular}{|c|c|c|c|c|c|c|}
\hline
intercept & $k$ & exchange & balance & coll. act. &  str. hole & $R^2$\\ 
\hline
$0.669^{***}$ & $0.0222^{***}$ & $0.691^{***}$ & $2.29^{***}$ & $3.15^{***}$ & $-1.34^{***}$ & 0.264 \\ % all
\hline
$[0.0414]$ & $[0.000362]$ & $[0.0571]$ & $[0.110]$ & $[0.0987]$ & $[0.126]$ & \\ 
\hline
$0.859^{***}$ & $0.0223^{***}$ & & $2.44^{***}$ & $3.23^{***}$ & $-1.31^{***}$ & 0.257 \\ % exchange
\hline
$[0.0385]$ & $[0.000364]$ & & $[0.110]$ & $[0.0989]$ & $[0.126]$ & \\ 
\hline
$0.738^{***}$ & $0.0228^{***}$ & $0.817^{***}$ & & $3.34^{***}$ & $-1.57^{***}$ & 0.245 \\ % balance
\hdashline
$[0.0418]$ & $[0.000365]$ & $[0.0575]$ & & $[0.0995]$ & $[0.127]$ & \\ 
\hline
$1.67^{***}$ & $0.0221^{***}$ & $0.814^{***}$ & $2.62^{***}$ & & $0.140^{***}$ & 0.219 \\ % col
\hdashline
$[0.0278]$ & $[0.000373]$ & $[0.0587]$ & $[0.113]$ & & $[0.121]$ & \\ 
\hline
$0.713^{***}$ & $0.0223^{***}$ & $0.675^{***}$ & $2.39^{***}$ & $2.77^{***}$ & & 0.259 \\ % hole
\hdashline
$[0.0413]$ & $[0.000363]$ & $[0.0573]$ & $[0.110]$ & $[0.0920]$ & & \\ 
\hline
$0.921^{***}$ & & $0.760^{***}$ & $2.85^{***}$ & $3.13^{***}$ & $-1.67^{***}$ & 0.0977 \\ % k
\hdashline
$[0.0456]$ & & $[0.0632]$ & $[0.122]$ & $[0.109]$ & $[0.139]$ & \\ 
\hline
$1.13^{***}$ & & & $3.01^{***}$ & $3.21^{***}$ & $-1.63^{***}$ & 0.0898 \\ % k a1
\hdashline
$[0.0423]$ & & & $[0.121]$ & $[0.109]$ & $[0.140]$ & \\ 
\hline
$1.02^{***}$ & & $0.920^{***}$ & & $3.36^{***}$ & $-1.96^{***}$ & 0.0680 \\ % k a2
\hdashline
$[0.0461]$ & & $[0.0639]$ & & $[0.111]$ & $[0.141]$ & \\ 
\hline
$1.91^{***}$ & & $0.882^{***}$ & $3.17^{***}$ & & $-0.19^{NS}$ & 0.0532 \\ % k c
\hdashline
$[0.0303]$ & & $[0.0646]$ & $[0.124]$ & & $[0.133]$ & \\ 
\hline
$0.978^{***}$ & & $0.741^{***}$ & $2.98^{***}$ & $2.64^{***}$ & & 0.0899 \\ % k a5
\hdashline
$[0.0455]$ & & $[0.0635]$ & $[0.122]$ & $[0.102]$ & & \\ 
\hline
\end{tabular}
\begin{flushleft}
\end{flushleft}
\label{tab:guest2}
\end{table}

\begin{table}[!ht]
\caption{
	\bf{Multilinear regression for the LJ network.} $M_i=c_0+c_1 a^{exc}_i +c_2 a^{bal}_i +c_3 a^{ca}_i + c_4 a^{sh}_i + c_5 k_i.$
}
\begin{tabular}{|c|c|c|c|c|c|c|}
\hline
intercept & $k$ & exchange & balance & coll. act. &  str. hole & $R^2$\\ 
\hline
$1.24^{***}$ & $0.00305^{***}$ & $1.30^{**}$ & $4.47^{***}$ & $0.804^{***}$ & $1.46^{***}$ & 0.195 \\ % all
\hdashline
$[0.0207]$ & $[5.76e$-5$]$ & $[0.111]$ & $[0.0513]$ & $[0.0545]$ & $[0.0549]$ & \\ 
\hline
$1.26^{***}$ & $0.00305^{***}$ & & $4.56^{***}$ & $0.787^{***}$ & $1.48^{***}$ & 0.194 \\ % exchange
\hdashline
$[0.0206]$ & $[5.76e$-5$]$ & & $[0.0507]$ & $[0.0545]$ & $[0.0549]$ & \\ 
\hline
$1.18^{***}$ & $0.00366^{***}$ & $2.88^{***}$ & & $2.52^{***}$ & $0.638^{***}$ & 0.103 \\ % balance
\hdashline
$[0.0219]$ & $[6.04e$-5$]$ & $[0.116]$ & & $[0.0537]$ & $[0.0571]$ & \\ 
\hline
$1.48^{***}$ & $0.00301^{***}$ & $1.26^{***}$ & $4.74^{***}$ & & $1.75^{***}$ & 0.193 \\ % col
\hdashline
$[0.0129]$ & $[5.76e$-5$]$ & $[0.111]$ & $[0.0479]$ & & $[0.0512]$ & \\ 
\hline
$1.23^{***}$ & $0.00304^{***}$ & $1.38^{***}$ & $4.23^{***}$ & $1.33^{***}$ & & 0.186 \\ % hole
\hdashline
$[0.0208]$ & $[5.79e$-5$]$ & $[0.112]$ & $[0.0508]$ & $[0.0510]$ & & \\ 
\hline
$1.32^{***}$ & & $1.32^{***}$ & $4.80^{***}$ & $0.671^{***}$ & $1.43^{***}$ & 0.160 \\ % k
\hdashline
$[0.0211]$ & & $[0.113]$ & $[0.0520]$ & $[0.0556]$ & $[0.0560]$ & \\ 
\hline
$1.35^{***}$ & & & $4.89^{***}$ & $0.6534{***}$ & $1.45^{***}$ & 0.159 \\ % k a1
\hdashline
$[0.0210]$ & & & $[0.0514]$ & $[0.0556]$ & $[0.0561]$ & \\ 
\hline
$1.27^{***}$ & & $3.03^{***}$ & & $2.51^{***}$ & $0.531^{***}$ & 0.0505 \\ % k a2
\hdashline
$[0.0224]$ & & $[0.119]$ & & $[0.0551]$ & $[0.0587]$ & \\ 
\hline
$1.52^{***}$ & & $1.28^{**}$ & $5.02^{***}$ & & $1.68^{***}$ & 0.158 \\ % k c
\hdashline
$[0.0132]$ & & $[0.113]$ & $[0.0486]$ & & $[0.0523]$ & \\ 
\hline
$1.31^{***}$ & & $1.39^{**}$ & $4.56^{***}$ & $1.19^{***}$ & & 0.152 \\ % k a5
\hdashline
$[0.0212]$ & & $[0.114]$ & $[0.0515]$ & $[0.0520]$ & & \\ 
\hline
\end{tabular}
\begin{flushleft}
\end{flushleft}
\label{tab:guest2}
\end{table}

\begin{table}[!ht]
\caption{
	\bf{Multilinear regression for the PRO network.} $M_i=c_0+c_1 a^{exc}_i +c_2 a^{bal}_i +c_3 a^{ca}_i + c_4 a^{sh}_i + c_5 k_i.$
}
\begin{tabular}{|c|c|c|c|c|}
\hline
intercept & $k$ & coll. act. &  str. hole & $R^2$\\ 
\hline
$-0.405^{***}$ & $0.137^{***}$ & $3.83^{***}$ & $3.67^{***}$ & 0.625 \\ % all
\hline
\hdashline
$[0.0382]$ & $[0.00128]$ & $[0.0810]$ & $[0.134]$ & \\ 
\hline
$-0.412^{***}$ & $0.139^{***}$ &  & $4.53^{***}$ & 0.594 \\ % a4
\hdashline
$[0.0398]$ & $[0.00133]$ & & $[0.0800]$ & \\ 
\hline
$1.07^{***}$ & $0.131^{***}$ & $5.65^{***}$ & & 0.533 \\ % a5
\hdashline
$[0.0246]$ & $[0.00142]$ & $[0.142]$ & & \\ 
\hline
$0.891^{***}$ & & $3.10^{***}$ & $4.50^{***}$ & 0.154 \\ % k
\hdashline
$[0.0545]$ & & $[0.121]$ & $[0.201]$ & \\ 
\hline
$2.05^{***}$ & & & $6.07^{***}$ & 0.093 \\ % k a4
\hdashline
$[0.0309]$ & & & $[0.198]$ & \\ 
\hline
$0.906^{***}$ & & $3.93^{***}$ & & 0.108 \\ % k a5
\hdashline
$[0.0559]$ & & $[0.118]$ & & \\ 
\hline
\end{tabular}
\begin{flushleft}
\end{flushleft}
\label{tab:guest2}
\end{table}

%\section*{Figures}
%SFig1
%\begin{figure}
%\begin{center}
%\includegraphics[width=\linewidth]{SFIG1.pdf}
%\caption{
%{\bf Scatter plots of the epidemic size as a function of the degree of node for the QXG, POK, and LJ networks.} 
%Color of symbols of each column represents the extent of social mechanisms for the 
%exchange, balance, collective action, and structural hole.
%The black line represents the average epidemic size for 
%the randomized networks conserving the degree distribution, obtained analytically.
%The nodes with high balance tendency locate in the high influential region.
%And, the collective action mechanism indicates the strong gradation horizontally.
%}
%\label{Figure_label}
%\end{center}
%\end{figure}


\begin{thebibliography}{99}
\bibitem{cohen} Cohen R, Havlin S, ben-Avraham D, Efficient Immunization Strategies for computer networks and populations. Phys. Rev. Lett. 2003; 91: 247901.
\bibitem{chen} Chen Y, Paul G, Havlin S, Liljeros F, Stanley HE, Finding a better immunization strategy. Phys. Rev. Lett. 2008; 101: 058701.
\bibitem{holme_immun} Holme P, Efficient local strategies for vaccination and network attack. Europhys. Lett. (EPL) 2004; 68: 908.
\bibitem{domingos} Domingos P, Richardson M, Mining the network value of customers. Seventh International Conference on Knowledge Discovery and Data Mining 2001.
\bibitem{domingos2} Domingos P, Richardson M, Mining knowledge-sharing sites for viral marketing. Eighth Intl. Conf. on Knowledge Discovery and Data Mining 2002.
\bibitem{albert} Albert R, Jeong H, Barab\'asi AL, Error and attack tolerance of complex networks. Nature 2000; 406: 378.
\bibitem{holme_attack} Holme P, Kim BJ, Yoon CN, Han SK, Attack vulnerability of complex networks. Phys. Rev. E 2002; 65: 056109.
\bibitem{kitsak} Kitsak M, Gallos LK, Havlin S, Liljeros F, Muchnik L, Stanley HE, Makse HA, Identification of influential spreaders in complex networks. Nat. Phys. 2010; 6: 888.
\bibitem{castellano} Castellano C, Pastor-Satorras R, Competing activation mechanisms in epidemics on networks. Sci. Rep. 2012; 2: 371.
\bibitem{pei} Pei S, Muchnik L, Andrade, Jr. JS, Zheng Z, Makse HA, Searching for superspreaders of information in real-world social media. Sci. Rep. 2014; 4: 5547.
\bibitem{kempe} Kempe D, Kleinberg J, Tardos E, Maximizing the spread of influence through a social network. Proc. 9th ACM SIGKDD Intl. Conf. on Knowledge Discovery and Data Mining 2003; 137-143.
\bibitem{kempe2} Kempe D, Kleinberg J, Tardos E, Influential nodes in a diffusion model for social networks. Proc. 32nd International Colloquium on Automata, Languages and Programming, ICALP 2005.
\bibitem{havlin} Cohen R, Erez K, Ben-Avraham D, Havlin S, Breakdown of the Internet under intentional attack, Phys. Rev. Lett. 2001; 86; 3682.
\bibitem{dorogovtsev} Dorogovtsev SN, Goltsev AV, Mendes JFF, K-core organization of complex networks. Phys. Rev. Lett. 2006; 96: 040601.
\bibitem{carmi} Carmi S, Havlin S, Kirkpatrick S, Shavitt Y, Shir E, A model of internet topology using k-shell decomposition. Proc. Nat. Acad. Sci. 2007; 104: 11150.
\bibitem{freeman} Freeman LC, Centrality in social networks: Conceptual clarification. Soc. Netw. 1979; 1: 215-239.
\bibitem{page} Brin S, Page L, The anatomy of a large-scale hypertextual web search engine. Comput. Networks ISDN 1998; 30: 107-117.
\bibitem{pei_makse} Pei S, Makse HA, Spreading dynamics in complex networks. J. Stat. Mech. 2013; P12002.
\bibitem{lee} Lee S, Kim P, Jeong H, Statistical properties of sampled networks. Phys. Rev. E 2006; 73: 016102.
\bibitem{contractor} Contractor NS, Wasserman S, Faust K, Testing multitheoretical, multilevel hypotheses about organizational networks: An analytic framework and empirical example, Acad. Manag. Rev. 2006; 31: 681.
\bibitem{monge} Monge PR, Contractor NS, Theories of communication networks. New York: Oxford University Press; 2003.
\bibitem{gallos} Gallos LK, Havlin S, Liljeros F, Makse HA, How people interact in evolving online affiliation networks. Phys. Rev. X 2012; 2: 031014.
\bibitem{barabasi} Barab\'asi AL, Albert R, Emergence of scaling in random networks. Science 1999; 286: 509.
\bibitem{ahn} Ahn Y, Bagrow JP, Lehmann S, Link communities reveal multiscale complexity in networks. Nature 2010; 466: 761-764.
\bibitem{fridlund} Fridlund V, Stenqvist K, Nordvik MK, Condom use: The discrepancy between practice and behavioral expectation. Scandinavian Journal of Public Health. 2014; 42: 759-765.
%\bibitem{lu_qx} Lu X, Bengtsson L, Britton T, Camitz M, Kim BJ, Thorson A, Liljeros F, The sensitivity of respondent-driven sampling. Journal of the Royal Statistical Society; Series A (Statistics in society) 2012; 175: 191-216.
\bibitem{holme_pok} Holme P, Edling CR, Liljeros F, Structure and time evolution of an Internet dating community. Soc. Netw. 2004; 26: 155.
\bibitem{rocha_pro} Rocha LEC, Liljeros F, Holme P, Information dynamics shape the sexual networks of Internet-mediated prostitution. Proc. Natl. Acad. Sci. 2010; 107: 5706.
\bibitem{clauset} Clauset A, Newman MEJ, Moore C, Finding community structure in very large networks. Phys. Rev. E 2004; 70. 066111.
\bibitem{anderson_may} Anderson RM, May RMC, Infectious diseases of humans. Oxford: Oxford University Press; 1991.
\bibitem{newman_epi} Newman MEJ, The spread of epidemic disease on networks, Phys. Rev. E 2002; 66: 016128.
\bibitem{pei_plos} Pei S, Muchnik L, Tang S, Zheng Z, Makse HA, Exploring the complex pattern of information spreading in online blog communities. arXiv:1504.00495.
\bibitem{gujarati} Gujarati D, Porter D, Basic Econometrics. 5th Ed. New York: McGraw-Hill; 1991.
\bibitem{pastor} Pastor-Satorras R, Vespignani A, Epidemic spreading in scale-free networks. Phys. Rev. Lett. 2006; 86: 3200.
\bibitem{granovetter} Granovetter MS, The strength of weak ties. American journal of sociology. 1973; 1360-1380.
\bibitem{montgomery} Montgomery JD, Job search and network composition: Implications of the strength of weak ties hypothesis. American Sociological Review 1992; 57: 586.
\bibitem{onnela} Onnela JP, Saram\"aki J, Hyv\"onen J, Szab\'o G, Lazer D, Kaski K, Kert\'esz J, Barab\'asi AL, Structure and tie strength in mobile communication networks. Proc. Nat. Acad. Sci. 2007; 104: 7332-7238.
\bibitem{sigman} Gallos LK, Makse HA, Sigman M, A small world of weak ties provides optimal global integration of self-similar modules in functional brain networks. Proc. Nat. Acad. Sci. 2012; 109: 2825.
\bibitem{masuda} Masuda N, Immunization of networks with community structure. New J. of Phys. 2009; 11: 123018.
\bibitem{modularity} Newman MEJ, Modularity and community structure in networks. Proc. Natl. Acad. Sci. 2006; 103; 8577-8696.
\end{thebibliography}

\section*{References}
\begin{thebibliography}{}
\bibitem{ahn} Ahn Y, Bagrow JP, Lehmann S, (2010) Link communities reveal multiscale complexity in networks, Nature {\bf 466}, 761-764.
\bibitem{newman_gf} Newman MEJ, Strogatz SH, Watts DJ, (2001) Random graphs with arbitrary degree distributions and their applications, Phys. Rev. E {\bf 64}, 041902.
\bibitem{newman_epi} Newman MEJ, (2002) The spread of epidemic disease on networks, Phys. Rev. E {\bf 66}, 016128.
\end{thebibliography}
\end{document}